\documentclass[draft]{agujournal2019}
\usepackage[T1]{fontenc}
\usepackage{url} 
\usepackage{lineno}
\usepackage[finalnew]{trackchanges} 
\usepackage{soul}
\usepackage{amsmath}
\usepackage{amssymb}
\usepackage{fix-cm}
\nolinenumbers

\newcommand{\ssr}{    {Space Sci. Rev.}}

\draftfalse

\journalname{JGR: Space Physics}

\begin{document}

%
%


\title{Plasma refilling of the lunar wake: plasma-vacuum interactions, electrostatic shocks, and electromagnetic instabilities}

%
%




\authors{Xin An\affil{1}, Vassilis Angelopoulos\affil{1}, Terry Z. Liu\affil{1}, Anton Artemyev\affil{1}, Andrew R. Poppe\affil{2}, Donglai Ma\affil{1}}


\affiliation{1}{Department of Earth, Planetary, and Space Sciences, University of California, Los Angeles, CA, 90095, USA}
\affiliation{2}{Space Sciences Laboratory, University of California, Berkeley, Berkeley, CA, 94720, USA}




\correspondingauthor{Xin An}{phyax@ucla.edu; https://sites.google.com/view/xin-an-physics}



\begin{keypoints}
\item Our particle-in-cell simulation validates plasma-vacuum interaction theory in the context of lunar wake refilling.
\item Counter-streaming supersonic ion beams form electrostatic shocks in the central wake, producing flattop electron distributions.
\item Temperature-anisotropic ion beams in the wake can generate ion-scale electromagnetic instabilities.
\end{keypoints}

%
%

%
%


\begin{abstract}
A plasma void forms downstream of the Moon when the solar wind impacts the lunar surface. This void gradually refills as the solar wind passes by, forming the lunar wake. We investigate this refilling process using a fully kinetic particle-in-cell (PIC) simulation. The early stage of refilling follows plasma-vacuum interaction theory, characterized by exponential decay of plasma density into the wake, along with ion acceleration and cooling in the expansion direction. Our PIC simulation confirms these theoretical predictions. In the next stage of the refilling process, the counter-streaming supersonic ion beams collide, generating Debye-scale electrostatic shocks at the wake's center. These shocks decelerate and thermalize the ion beams while heating electrons into flat-top velocity distributions along magnetic field lines. Additionally, fast magnetosonic waves undergo convective growth via anomalous cyclotron resonance as they co-propagate with temperature-anisotropic ion beams toward the wake's center. Electromagnetic ion cyclotron waves may also be excited through normal cyclotron resonance, counter-propagating with these anisotropic ion beams. Our findings provide new insights into the kinetic aspects of lunar wake refilling and may enhance interpretation of spacecraft observations.
\end{abstract}

\section*{Plain Language Summary}
When the solar wind (a stream of electrically charged particles from the Sun) flows around the Moon, it creates an empty region behind it called the lunar wake. This happens because the Moon absorbs the solar wind particles that hit its surface. Our research investigates how this empty space gradually refills with particles as the solar wind continues flowing past the Moon. Using advanced computer simulations that track individual particles, we studied this refilling process in detail. In the early stages, particles from the surrounding solar wind flow into the empty region, becoming less dense the deeper they get into the wake. The fastest particles extend farthest into the wake and appear cooler as they move inward. When particles flowing in from opposite sides meet in the center of the wake, they create short-scale shock waves. They also generate different types of electromagnetic waves. These waves may be detectable by spacecraft orbiting the Moon. Our findings help better understand the complex behavior of space plasmas around airless bodies like the Moon and could improve the interpretation of past and future spacecraft measurements.

%
%

%


%
%
%
%

\section{Introduction}
The Moon, as an airless, nonconductive body lacking a global intrinsic magnetic field, does not generate global effects on incoming plasma flows and magnetic fields through mass and momentum loading. When supersonic solar wind plasma impacts the dayside lunar surface, it creates a void downstream that gradually refills, forming an extended tail known as the lunar wake. The theoretical framework for plasma-vacuum interactions developed by \citeA{gurevich1966self} and expanded by others \cite<e.g.,>[]{crow1975expansion,denavit1979collisionless,mora2003plasma} provides a model for wake refilling along magnetic field lines. This theory demonstrates that plasma expansion into a vacuum generates both a rarefaction wave propagating into the ambient solar wind and a charge-separation electric field that accelerates ions into the vacuum region. Broadband electrostatic turbulence has been observed at the wake's center where counter-streaming plasmas collide \cite{halekas2014first}—a phenomenon predicted by kinetic simulations, although the precise excitation mechanisms involving counter-streaming electron or ion beams remain elusive \cite{farrell1998simple,birch2001detailed}. Additionally, the ion temperature parallel to the magnetic field decreases exponentially toward the central wake, while the perpendicular ion temperature remains relatively constant. The resultant ion anisotropy has been hypothesized to trigger ion cyclotron instabilities \cite{gurevich1966self}.


The lunar wake was first detected by Explorer 35 and the Apollo subsatellites \cite<see reviews by>[]{ness1972interaction,schubert1974observations}. Early investigations primarily focused on magnetic field measurements \cite{colburn1967diamagnetic,ness1967early,ness1968perturbations}, as particle detection capabilities were insufficient to fully characterize velocity distributions \cite{lyon1967explorer,siscoe1969experimental}. In the 1990s, the Wind spacecraft conducted multiple lunar flybys, providing new magnetic field measurements \cite{owen1996lunar} along with enhanced capabilities for observing particle distribution functions \cite{bosqued1996moon,ogilvie1996observations,clack2004wind} and plasma waves both within and upstream of the wake \cite{farrell1996upstream,kellogg1996observations,bale1997evidence}.\add{ Using electron and magnetic field measurements, the Lunar Prospector spacecraft revealed the wake morphology at altitudes of $\lesssim 100$\,km, including the wake potential drop and diamagnetic current system} \cite{halekas2005electrons}. During the 2000s, three missions—the Japanese Kaguya, Chinese Chang'E, and Indian Chandrayaan—contributed new measurements of the low-altitude wake, demonstrating the importance of finite ion gyroradius effects \cite{nishino2009pairwise,nishino2009solar,wang2010acceleration,dhanya2016characteristics} and revealing plasma wave activities \cite{hashimoto2010electrostatic} in this region. Since 2011, the twin ARTEMIS spacecraft \cite{angelopoulos2014artemis} have routinely traversed the wake at varying distances from the Moon, delivering comprehensive measurements of DC magnetic and electric fields, proton and electron distributions, and wave fields. These accumulated observations have provided significant new insights into plasma kinetics \cite{halekas2014first,halekas2014effects,halekas2015moon,tao2012kinetic} and the global structure of the lunar wake \cite{zhang2012outward,zhang2014three,xu2019mapping,xu2020reflected}.

Multidimensional hybrid simulations (where ions are treated as particles and electrons as a fluid) provide a global view of the lunar wake, capturing magnetic field structures \cite{wang20113d}, current systems \cite{fatemi2013lunar}, and ion phase space structures with related wave instabilities \cite{travnivcek2005structure,fatemi2012effects}. The missing electron-scale kinetic processes can be resolved through fully kinetic simulations (where both ions and electrons are treated as particles), which offer a microscopic view of Debye-scale electrostatic waves and structures in the wake \cite{farrell1998simple,birch2001detailed,birch2001particle,birch2002two}. To manage computational costs, these early simulations assumed a small Moon size ($\leq 256$ Debye lengths) and reduced ion-to-electron mass ratio ($\leq 25$). With the advent of petascale ($10^{15}$ floating point operations per second, or $10^{15}$ FLOPS) and exascale ($10^{18}$ FLOPS) computing, this study revisits the wake refilling problem from a fully kinetic perspective using more realistic physical parameters.\remove{ As a first step, in this paper we aim to validate our simulation against established plasma-vacuum interaction theory, clarify the controversial excitation mechanisms of electrostatic instabilities in the central wake (whether by electron or ion beams), and investigate the possible excitation of ion-scale electromagnetic instabilities.}

\add{In this study, we investigate three specific processes during plasma refilling of the lunar wake: (1) plasma-vacuum interactions, (2) electrostatic shocks, and (3) electromagnetic instabilities.}

\add{First, plasma expansion into a vacuum is well described by self-similar solutions of different plasma moments (density, flow velocity, and temperature) and ion front motions in both fluid and kinetic theories }\cite{gurevich1966self,crow1975expansion,denavit1979collisionless,mora2003plasma}\add{. These solutions provide an excellent benchmark to validate our particle-in-cell (PIC) simulations. Our new contribution to understanding plasma-vacuum interactions is a more accurate equation of state for electrons, $T_{e,x}/n_e = \text{constant}$ (where $T_{e,x}$ is the electron temperature in the expansion direction and $n_e$ is the plasma density), which describes electron cooling caused by the ambipolar electric field beyond the adiabatic limit.}

\add{Second, while previous PIC simulations show certain signatures of electrostatic instabilities }\cite{farrell1998simple,birch2001detailed,birch2001particle,birch2002two}\add{, it remains controversial whether they are excited by electron or ion beams. Moreover, given that the characteristic spatial scales of these electrostatic solitary structures are tens of Debye lengths }\cite<e.g.,>[]{sun2022double}\add{, it is questionable whether the relatively small Moon sizes ($\leq 256$ Debye lengths) in previous simulations allow sufficient development of electrostatic instabilities into shocks and their subsequent long-timescale evolution. Our new contribution addresses this limitation by elucidating the formation of ion-acoustic-type electrostatic shocks by supersonic ion beams and quantifying the energy transfer from the kinetic energy of ion beams to the thermal energy of flat-top electron distributions. This analysis is enabled by the more realistic Moon size ($\sim 4.8 \times 10^4$ Debye lengths) and detailed energy budget calculations.}

\add{Third, ion beams naturally develop temperature anisotropies as they are accelerated into the lunar wake due to exponential cooling along the background magnetic field. It has been hypothesized that these anisotropic ion beams would excite ion-scale electromagnetic instabilities }\cite{gurevich1966self}\add{. While the Moon sizes ($\leq 256$ Debye lengths) in previous simulations }\cite{farrell1998simple,birch2001detailed,birch2001particle,birch2002two}\add{ are insufficient to accommodate these ion-scale waves, we validate this hypothesis by demonstrating the excitation of magnetosonic waves (and possibly electromagnetic ion cyclotron waves). This validation is enabled by our extended Moon size ($26.4$ ion inertial lengths) and awaits further confirmation from spacecraft observations.}

The remaining sections are organized as follows. Section \ref{sec:theory} briefly presents the theory of plasma-vacuum interactions that will be compared against our PIC simulation. Section \ref{sec:setup} introduces the computational setup of the PIC simulation. Section \ref{sec:plasma-vacuum} compares the PIC simulation results against theoretical predictions presented in Section \ref{sec:theory}. Section \ref{sec:electrostatic-shocks} examines the formation of electrostatic shocks and the associated energy conversion in the central wake. Section \ref{sec:electromagnetic-instabilities} investigates the ion-scale electromagnetic instabilities generated by anisotropic ion beams in the wake. Section \ref{sec:summary} summarizes the main findings of the present study.

\section{Theory}\label{sec:theory}
\begin{figure}[tphb]
    \centering
    \includegraphics[width=0.85\linewidth]{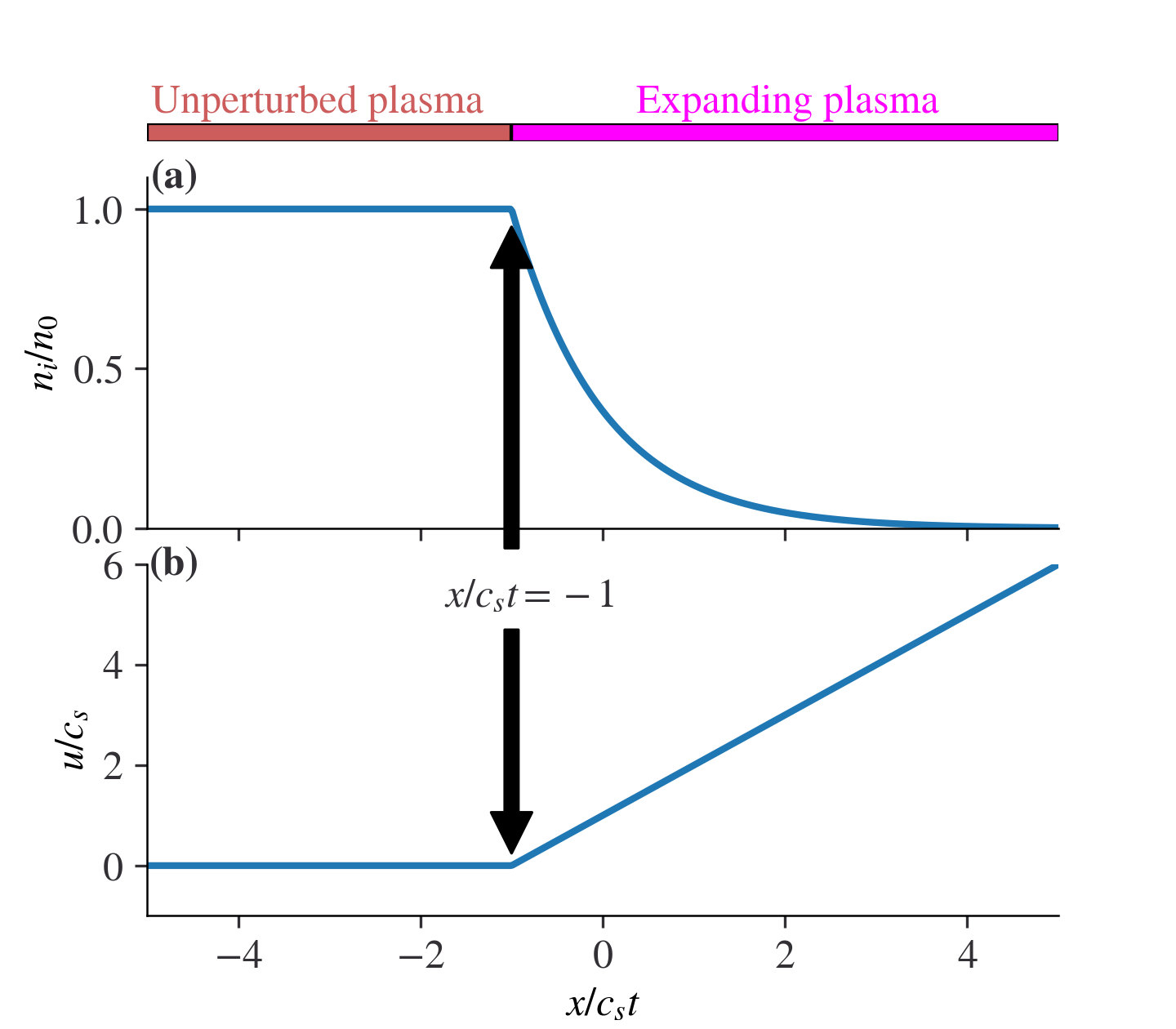}
    \caption{Spatial profiles of (a) plasma density and (b) flow velocity, shown in the self-similar coordinate $x/c_s t$. The arrows mark the rarefaction wave front at $x + c_s t = 0$, which propagates leftward.}
    \label{fig:fluid-self-sim}
\end{figure}

We recap the basic theory of plasma-vacuum interactions to help interpret our kinetic simulations. Consider a plasma initially occupying the half-space $x<0$ at time $t=0$, with a sharp density discontinuity: both ion and electron densities are $n_i = n_e = n_0$ for $x<0$ and zero for $x>0$. Due to their higher thermal velocity, electrons expand into the vacuum faster than ions, creating a charge-separation electric field. This field simultaneously accelerates ions outward while restraining the electron expansion. Since ion thermal motion plays a minor role, the spatiotemporal evolution of ion density $n_i$ and streaming velocity $u$ is governed by \cite<e.g.,>[]{crow1975expansion,mora2003plasma}:
\begin{linenomath*}
    \begin{align}
        \frac{\partial n_i}{\partial t} + u\frac{\partial n_i }{\partial x} = -n_i \frac{\partial u}{\partial x}, \label{eq:continuity}\\
        \frac{\partial u}{\partial t} + u\frac{\partial u}{\partial x} = -\frac{e}{m_i} \frac{\partial \Phi}{\partial x} . \label{eq:motion}
    \end{align}
\end{linenomath*}
Here, $\Phi$ represents the electric potential, $e$ is the elementary charge, and $m_i$ is the ion mass. For spatial scales larger than the local Debye length and time scales longer than the local ion plasma period, we can invoke the quasi-neutrality condition $n_i = n_e$. Assuming isothermal electrons ($T_e = \mathrm{constant}$) that follow a Boltzmann distribution $n_e = n_0 \exp(e \Phi / T_e)$, the electric field term in Equation \eqref{eq:motion} simplifies to $-c_s^2 \partial (\ln{n_i/n_0}) / \partial x$, where $c_s = \sqrt{T_e / m_i}$ is the ion acoustic speed. Note that the Boltzmann constant $k_B$ has been absorbed into the temperature $T_e$. Under the quasi-neutrality condition, Equations \eqref{eq:continuity} and \eqref{eq:motion} become scale-invariant (i.e., no longer containing any parameters with units of time or length), yielding a self-similar solution for $x + c_s t > 0$:
\begin{linenomath*}
    \begin{align}
        u &= c_s \left(\frac{x}{c_s t} + 1\right) , \label{eq:ss-vel} \\
        n_i &= n_0 \exp\left( -\frac{x}{c_s t} - 1 \right) , \label{eq:ss-den} \\
        E_{ss} &= \frac{T_e}{e c_s t} = \frac{T_e}{e \lambda_{D0}} \frac{1}{\omega_{pi} t} \label{eq:ss-efld} .
    \end{align}
\end{linenomath*}
Here, $\lambda_{D0} = \sqrt{T_e / (4 \pi n_0 e^2)} = c_s / \omega_{pi}$ is the Debye length of the unperturbed plasma and $\omega_{pi} = \sqrt{4 \pi n_0 e^2 / m_i}$ is the ion plasma frequency of the unperturbed plasma. Figure \ref{fig:fluid-self-sim} shows the solution in the self-similar coordinate: $x/(c_s t)$. The boundary between perturbed and unperturbed plasma regions is located at $u = 0$, yielding $x_r = -c_s t$. This indicates that a rarefaction wave propagates into the unperturbed plasma at the velocity $-c_s$. The self-similar electric field given by Equation \eqref{eq:ss-efld} can be interpreted as arising from two surface charge layers: a positive layer with density $\sigma = E_{ss} / 4 \pi$ at the rarefaction wave front $x = x_r$ and a negative layer with density $-\sigma = -E_{ss}/4\pi$ at the \change{expansion}{ion} front \cite{mora2003plasma}.\add{ The concept of the ion front is introduced as follows.}

Equation \eqref{eq:ss-vel} predicts the ion velocity increases without a limit as $x \to +\infty$. Physically an ion front exists where the local Debye length, $\lambda_D = \lambda_{D0} \sqrt{\exp[x/( c_s t) + 1]}$, becomes comparable to the density scale length, $c_s t$ \cite{pearlman1978maximum,mora2003plasma}. The location of the ion front is given by
\begin{linenomath*}
    \begin{align}
        \frac{x_f}{c_s t} + 1 = 2 \ln(\omega_{pi} t) . \label{eq:ss-xfront}
    \end{align}
\end{linenomath*}
This solution is physically meaningful only when the expansion time greatly exceeds the inverse ion plasma frequency ($t \omega_{pi} \gg 1$), or equivalently, when the expansion length scale greatly exceeds the initial Debye length ($c_s t \gg \lambda_{D0}$).

In the region beyond the ion front ($x > x_f$), the quasi-neutrality approximation breaks down. The ion front velocity can be obtained from Equations \eqref{eq:ss-xfront} and
\begin{linenomath*}
    \begin{align}
        u_f = c_s \left(\frac{x_f}{c_s t} + 1\right) = 2 c_s \ln(\omega_{pi} t) .
    \end{align}
\end{linenomath*}
The corresponding electric field at the ion front is:
\begin{linenomath*}
    \begin{align}\label{eq:efld-ifront}
        E_f = \frac{m_i}{e} \frac{\mathrm{d} u_f}{\mathrm{d} t} = \frac{2 T_e}{e \lambda_{D0}} \frac{1}{\omega_{pi} t} = 2 E_{ss} ,
    \end{align}
\end{linenomath*}
which is twice the magnitude of the self-similar electric field. A more rigorous derivation of this electric field can be obtained by solving the ion fluid equations coupled with Poisson's equation, rather than invoking the quasi-neutrality approximation \cite{mora2003plasma}.

Self-similar solutions to the ion Vlasov equation were first derived by \citeA{gurevich1966self}, who demonstrated exponential cooling parallel to the expansion direction. The evolution of the ion phase space density $f(x, v, t)$ is governed by
\begin{linenomath*}
    \begin{align}
        \frac{\partial f}{\partial t} + v \frac{\partial f}{\partial x} - \frac{e}{m_i} \frac{\partial \Phi}{\partial x} \frac{\partial f}{\partial v} = 0 .
    \end{align}
\end{linenomath*}
Under the assumptions of quasi-neutrality ($n_e = n_i = n_0 \int f \mathrm{d}v$) and Boltzmann-distributed electrons ($n_e = n_0 \exp(e \Phi / T_e)$), the electric potential can be expressed as $\Phi = (T_e / e) \ln (\int f \,\mathrm{d}v)$. This substitution makes the ion Vlasov equation scale-invariant. By introducing the similarity variable $\xi = x / t$ and assuming $f = f(\xi, v)$, we obtain
\begin{linenomath*}
    \begin{align}\label{eq:ss-vlasov}
    (v - \xi) \frac{\partial f}{\partial \xi} + c_s F \frac{\partial f}{\partial v} = 0 ,
    \end{align}
where $F$ is a dimensionless force given by
\begin{align}
    F(\xi) = - c_s \frac{\partial (\ln \int f\,\mathrm{d}v)}{\partial \xi} .
\end{align}
\end{linenomath*}
The force $F$ is related to the electric field through $E_x / F = T_e / (e c_s t) = T_e / (e \lambda_{D0} \omega_{pi} t)$, and exhibits asymptotic behavior where $F \to 0$ as $\xi \to -\infty$ and $F \to 1$ as $\xi \to +\infty$.

\begin{figure}[tphb]
    \centering
    \includegraphics[width=0.85\linewidth]{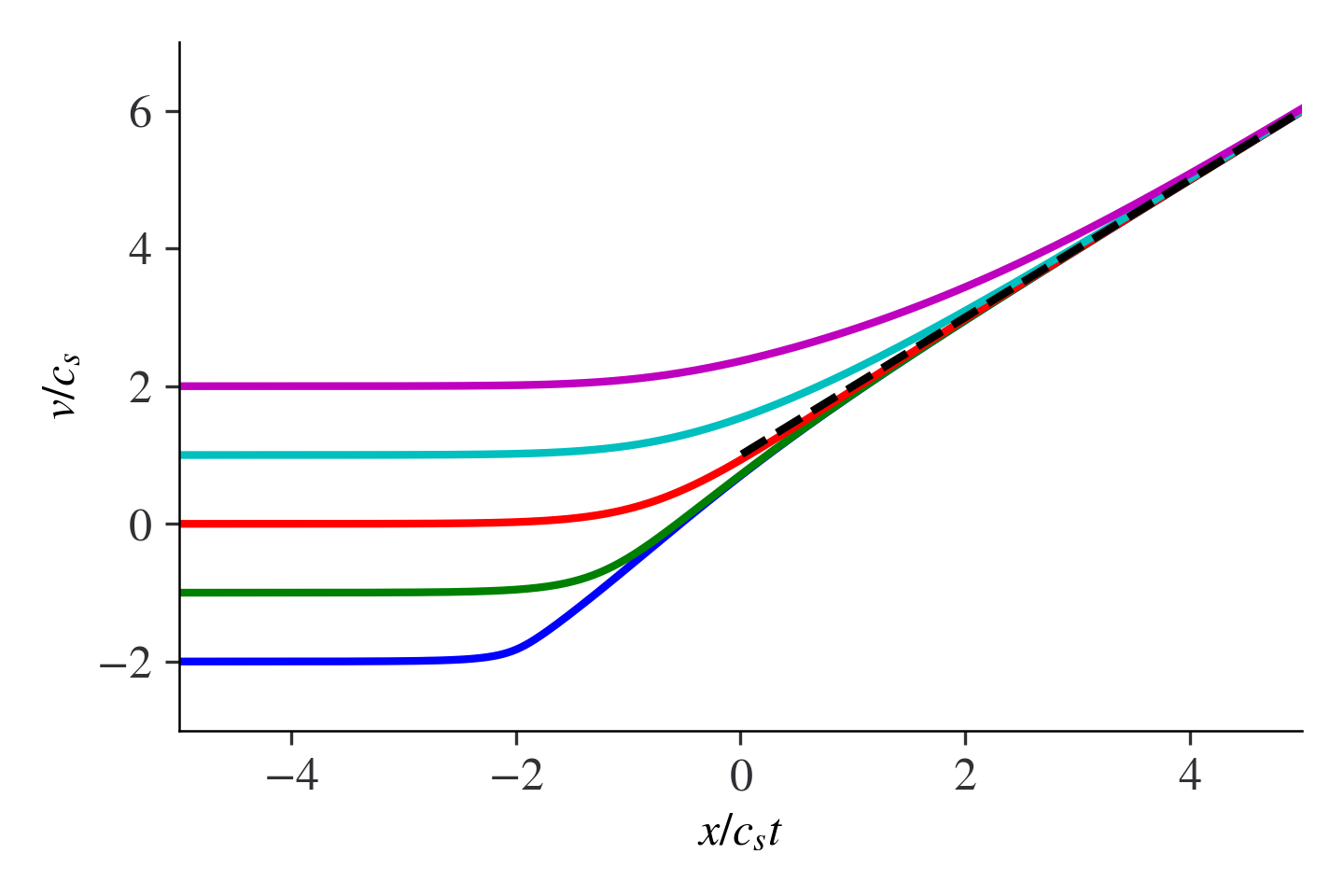}
    \caption{Sketch of ion motion characteristics in the phase space $(\xi, v)$. The dashed line represents the asymptote $v = \xi + c_s$.}
    \label{fig:kinetic-self-sim}
\end{figure}

The phase space density $f$ is conserved along the characteristics described by
\begin{linenomath*}
    \begin{align}
        \frac{\mathrm{d} v}{\mathrm{d} \xi} = \frac{c_s F}{v - \xi} . \label{eq:characteristics}
    \end{align}
\end{linenomath*}
Figure \ref{fig:kinetic-self-sim} illustrates these characteristics for several initial ion velocities at $\xi \to -\infty$. The characteristics exhibit two asymptotic behaviors: as $\xi \to -\infty$, $\mathrm{d} v / \mathrm{d} \xi = 0$, and as $\xi \to +\infty$, $\mathrm{d} v / \mathrm{d} \xi = c_s / (v - \xi)$. The latter case can be integrated to yield
\begin{linenomath*}
    \begin{align}
        v = \xi + c_s + A \exp(-v/c_s) ,
    \end{align}
\end{linenomath*}
\add{which asymptotically approaches the ion flow velocity in Equation }\eqref{eq:ss-vel}\add{, $v \to \xi + c_s$, for large $\xi$. The third term in this equation constitutes a higher-order correction that decreases exponentially with $\xi$. Since $A \exp(-v / c_s) \approx A \exp[-(\xi + c_s) / c_s] = A' \exp (-\xi / c_s)$, where $A' = A \exp(-1)$, both $A$ and $A'$ are integration constants encoding ion phase information. }\remove{where $A$ and $A'$ are integration constants. This solution asymptotically approaches $v = \xi + c_s$, which matches the flow velocity predicted by the fluid equations [Equation (3)]. }The ion thermal velocity \change{can be characterized by}{is determined from the velocity fluctuations about the mean flow:}
\begin{linenomath*}
    \begin{align}\label{eq:ion-temprature}
        v_{\mathrm{Ti}}^2 &= \langle (v - \langle v \rangle)^2 \rangle \nonumber \\
    & \approx \langle A'^2 \exp(-2\xi / c_s) \rangle \nonumber \\
    & = \langle A'^2 \rangle \exp(-2\xi / c_s) \nonumber \\
    &= C \cdot \exp(-2\xi / c_s) ,
    \end{align}
\end{linenomath*}
\add{where $\langle \cdot \rangle$ denotes the average over initial ion phases, and $C = \langle A'^2 \rangle$ is a constant. This equation }\change{demonstrating}{demonstrates} the exponential cooling of ions in the expansion direction.

\section{Computational setup}\label{sec:setup}
\begin{figure}[tphb]
    \centering
    \includegraphics[width=0.85\linewidth]{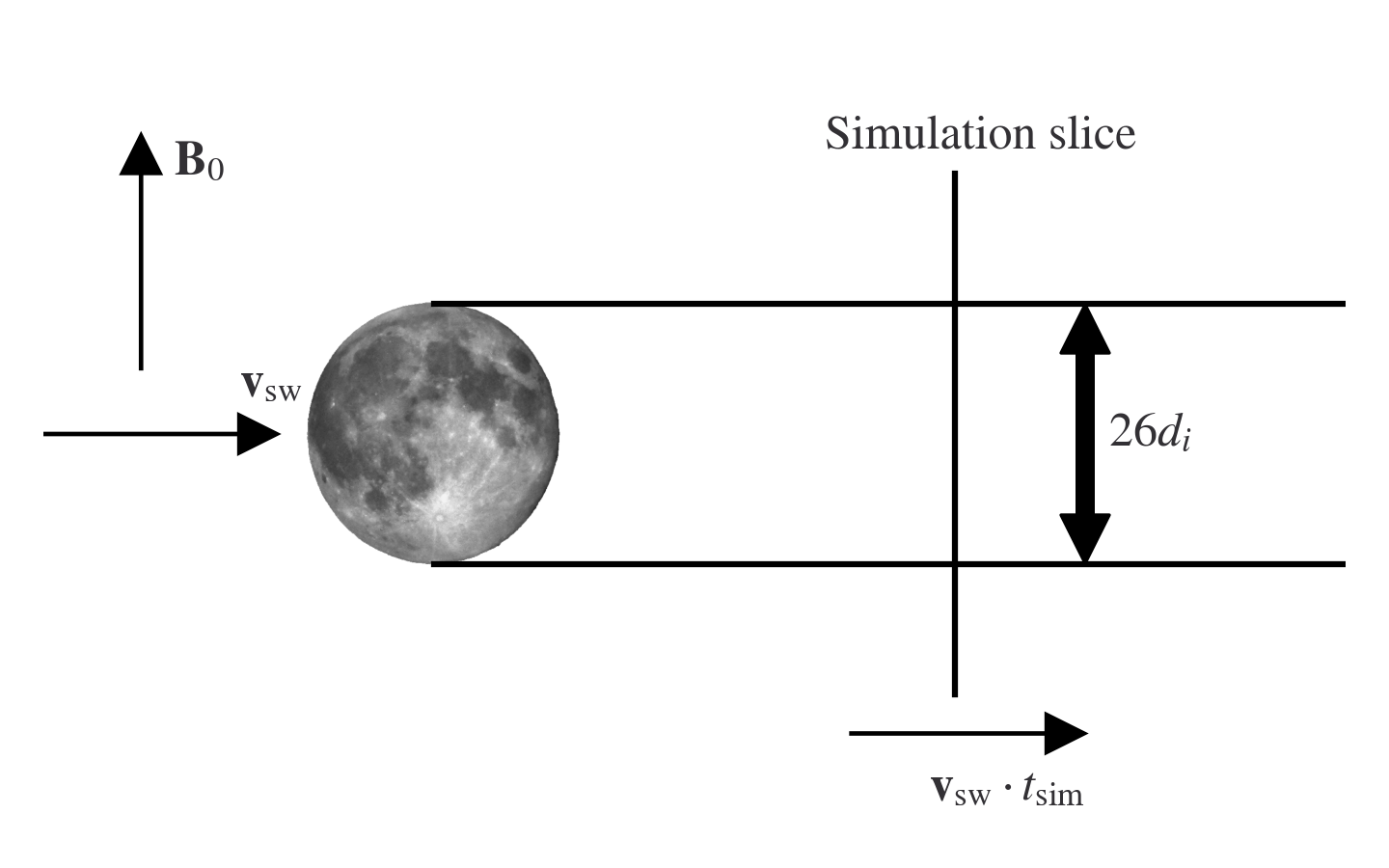}
    \caption{Schematic of the computational domain. The one-dimensional simulation slice is aligned parallel to the ambient magnetic field $\mathbf{B}_0$ and convects with the solar wind flow velocity. For scale reference, the lunar diameter corresponds to approximately $26\,d_i$, where $d_i$ is the ion inertial length in the solar wind.}
    \label{fig:setup}
\end{figure}

We use the electromagnetic, massively parallel Vector Particle-in-Cell (VPIC) code \cite{bird2021vpic,bowers2008ultrahigh,bowers20080,bowers2009advances} to simulate plasma refilling in the lunar wake. Figure \ref{fig:setup} illustrates our computational setup. We model a one-dimensional slice of plasma extending through the lunar shadow, as it is convected by the solar wind flow along the direction of the wake. The simulation's temporal evolution directly maps to a spatial evolution according to $r = v_{\mathrm{sw}} \cdot t_{\mathrm{sim}}$, where $r$ represents the radial distance from the Moon, $v_{\mathrm{sw}}$ is the solar wind velocity, and $t_{\mathrm{sim}}$ is the simulation time. This approach follows previous studies by \citeA{birch2001detailed,birch2001particle}. In this study, we align the solar wind magnetic field $\mathbf{B}_0$ parallel to our simulation slice. More complex configurations--such as non-orthogonal $\mathbf{v}_{\mathrm{sw}}$ and $\mathbf{B}_0$ orientations or plasma refilling dynamics perpendicular to $\mathbf{B}_0$-- would require at least two-dimensional configuration space simulations and computational cost reduction strategies, which we leave for future investigations.

To manage computational costs, we adopt a reduced ion-to-electron mass ratio $m_i / m_e = 100$. Table \ref{tab:sim_parameters} provides a comprehensive comparison of dimensionless parameters between the realistic ratio ($m_i/m_e = 1836$) and our reduced ratio ($m_i/m_e = 100$). By maintaining the electron temperature $T_e$ at realistic values, we preserve the ion acoustic velocity $c_s$, the characteristic velocity of plasma expansion into a vacuum. Although the reduced mass ratio artificially narrows the separation between ion and electron characteristic spatiotemporal scales, it preserves the essential velocity ordering: $v_{\mathrm{Te}} \gg c_s > v_{\mathrm{Ti}}$. Maintaining this hierarchy ensures that the fundamental physics governing wake refilling dynamics remains valid.


\begin{table}[tphb]
    \centering
    \begin{tabular}{||c|c|c|c|c|c|c||}
    \hline
    $m_i / m_e$ & $\omega_{pi}/\omega_{ci}$ & $\omega_{pe}/\omega_{pi}$ & $\omega_{pi}/\omega_{ce}$ & $c/v_\mathrm{A}$ & $c/v_{\mathrm{Te}}$ & $c/v_{\mathrm{Ti}}$ \\
    \hline
    $1836$ & $7955$ & $43$ & $4.3$ & $7955$ & $185$ & $9686$ \\
    $100$ & $7955$ & $10$ & $79.55$ & $7955$ & $185$ & $2261$ \\
    \hline\hline
    $m_i / m_e$  & $R_l/d_i$ & $d_i / \lambda_D$ & $d_i/d_e$ & $v_{\mathrm{Ti}}/c_s$ & $v_{\mathrm{Te}}/c_s$ & $c_s/v_{\mathrm{A}}$ \\
    \hline
    $1836$ & $13.2$ & $7923$ & $43$ & $0.82$ & $43.0$ & $1.0$ \\
    $100$ & $13.2$ & $1846$ & $10$ & $0.82$ & $10.0$ & $4.3$ \\
    \hline
    \end{tabular}
    \caption{Comparison of dimensionless parameters between reference values (using realistic ion-to-electron mass ratio) and reduced mass ratio values employed in lunar wake dynamics simulations. Parameters include lunar radius ($R_l$), electron and ion inertial lengths ($d_e$, $d_i$), thermal velocities ($v_{\mathrm{Ti}}$, $v_{\mathrm{Te}}$), Alfv\'en velocity ($v_{\mathrm{A}}$), and characteristic frequencies ($\omega_{pe}$, $\omega_{pi}$, $\omega_{ce}$, $\omega_{ci}$). Reference solar wind conditions with $m_i / m_e = 1836$ use: magnetic field $B_0 = 3$\,nT, plasma density $n_0 = 3\,\mathrm{cm}^{-3}$, temperatures $T_i = 10$\,eV and $T_e = 15$\,eV, and lunar radius $R_l = 1740$\,km.}
    \label{tab:sim_parameters}
\end{table}

Our computational domain spans $-L_x/2 \leq x \leq L_x/2$ with $L_x = 60\,d_i$, discretized using $N_x = 300,000$ grid points. This provides a spatial resolution of $\Delta x = 0.0002\,d_i = 0.002\,d_e = 1.6\,\lambda_D$. To satisfy the Courant condition $\Delta t < \Delta x / c$, we set the time step to $\Delta t = 0.00198\,\omega_{pe}^{-1}$. The initial density profile is defined as:
\begin{linenomath*}
    \begin{align}
        n(x) = \begin{cases}
        n_0, & R_l < \vert x \vert < L_x / 2 , \\
        0, & 0 \leq \vert x \vert \leq R_l ,
        \end{cases}
    \end{align}
\end{linenomath*}
where $n_0$ represents the reference density in the solar wind. The vacuum region at $\vert x \vert \leq R_l$ models the plasma void immediately downstream of the Moon. Each cell in the reference density region contains $400$ particles of each species (electrons and ions), resulting in a total of $1.36 \times 10^8$ particles initially. The initial velocity distribution for both species follows a Maxwellian profile
\begin{linenomath*}
    \begin{align}
        f_{\alpha}(x, v_x, v_y, v_z) \propto n(x) \exp\left(-\frac{v_x^2 + v_y^2 + v_z^2}{2 v_{\mathrm{T}\alpha}^2}\right) ,
    \end{align}
\end{linenomath*}
where $\alpha = i, e$ denotes ions and electrons, respectively. For particle boundary conditions, we implement open boundaries: particles crossing the domain boundaries are removed from the simulation, while new particles are injected at both boundaries by sampling from one-sided Maxwellian velocity distributions. For electromagnetic fields, we employ absorbing boundary conditions without image charges, with charge and current densities accumulated only over partial voxels at the boundaries.

\begin{figure}[tphb]
    \centering
    \includegraphics[width=0.85\linewidth]{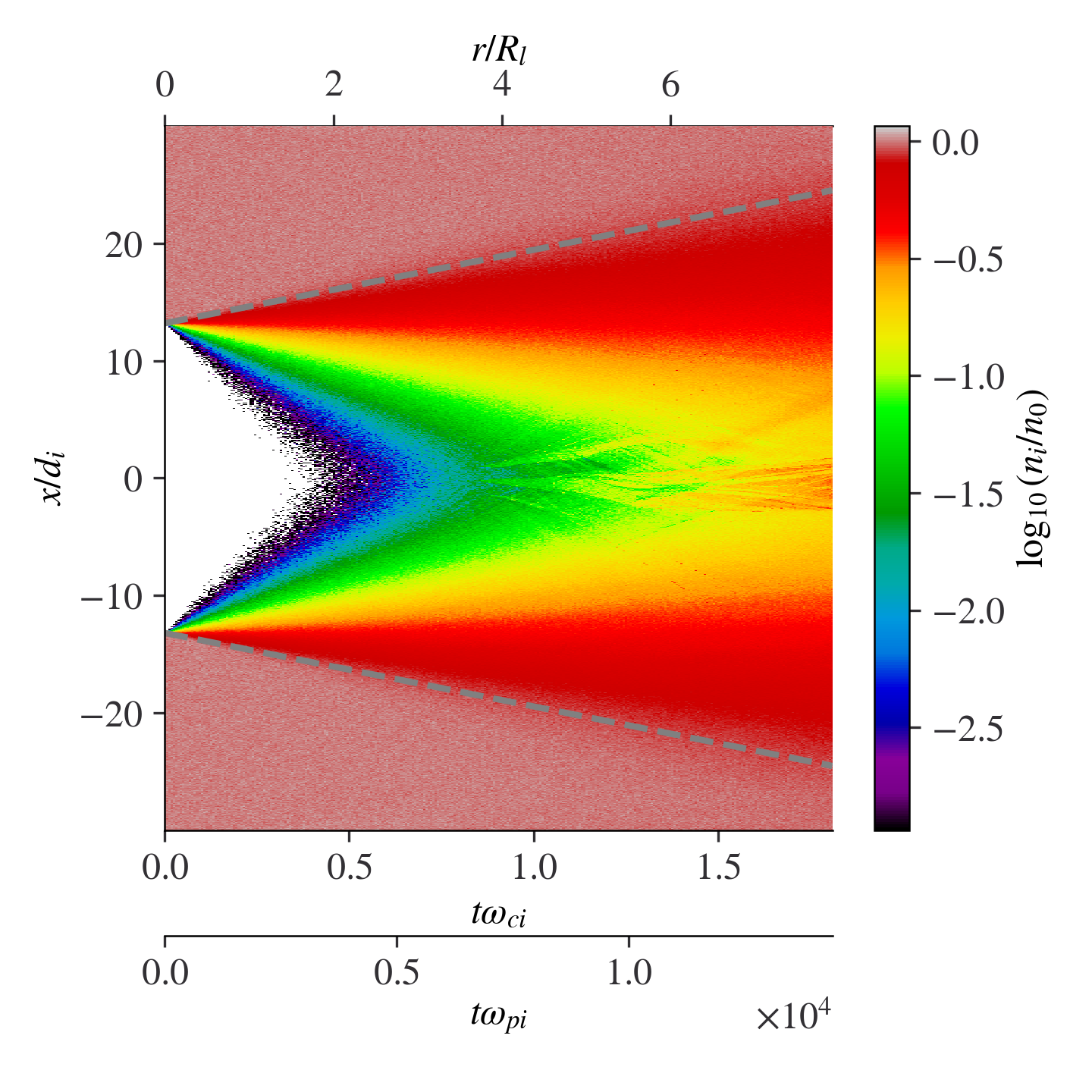}
    \caption{Spatiotemporal evolution of plasma density in the simulation. The dashed lines indicate the propagation characteristics of rarefaction waves, following the relation $(x\pm R_l) / (c_s t) \pm 1 = 0$.}
    \label{fig:tjoin-density}
\end{figure}

Figure \ref{fig:tjoin-density} illustrates the temporal evolution of plasma density across the 1D simulation domain. It captures two key phenomena: (1) the propagation of rarefaction waves from boundaries at $x = \pm R_l$ into the ambient solar wind, and (2) the complex density structures formed in the wake's center due to counter-streaming plasmas.

By mapping the time axis to spatial distance via $v_{sw} t$, this figure provides a first glimpse of the lunar wake structure.\add{ The mapping from simulation time to normalized radial distance is given by:}
\begin{linenomath*}
    \begin{align}
    \frac{r}{R_l} = \frac{v_{\mathrm{sw}} t}{R_l} = \left(\frac{v_{\mathrm{sw}}}{v_{\mathrm{A}}} \frac{v_{\mathrm{A}}}{c_s} \right) \left( \frac{c_s}{\omega_{pi} R_l} t \omega_{pi} \right) = \left(M_{\mathrm{A}} \sqrt{\frac{2}{\beta_i} \frac{T_i}{T_e}}\right) \left(\frac{\lambda_D}{R_l} t \omega_{pi}\right) ,
    \end{align}
\end{linenomath*}
\add{where we have used the Alfv\'en Mach number $M_{\mathrm{A}} = v_{\mathrm{sw}} / v_{\mathrm{A}}$, $v_{\mathrm{A}} / c_s = \sqrt{(2/\beta_i)(T_i / T_e)}$, and $\lambda_D = c_s / \omega_{pi}$. We apply reference solar wind conditions from }\citeA{fatemi2013lunar}\add{: $M_{\mathrm{A}} = 7.9$, $\beta_i = 0.6$, and $T_i / T_e = 0.86$. With our simulation parameter $R_l / \lambda_D = 2.4 \times 10^4$, this yields}
\begin{linenomath*}
    \begin{align}
        \frac{r}{R_l} = 5.5 \times 10^{-4} t \omega_{pi} .
    \end{align}
\end{linenomath*}
\add{At simulation times $t \omega_{pi} = 7500$ and $t \omega_{pi} = 14400$, this yields $r/R_l = 4.1$ and $r/R_l = 7.9$, respectively. These values are of the same order of magnitude as those in }\citeA{fatemi2013lunar}\add{, confirming that our simulation captures the relevant physics at the appropriate scale.}

Computationally, the simulation required approximately $240$\,hours on $1024$ processors (AMD EPYC 7763 Milan) to reach $t \omega_{pi} = 14400$ on the supercomputer Derecho \cite<a 19.87-petaflops system; see>[]{derecho}.

\section{Plasma-vacuum interactions}\label{sec:plasma-vacuum}
\subsection{Self-similar solutions}
\begin{figure}[tphb]
    \centering
    \includegraphics[width=0.75\linewidth]{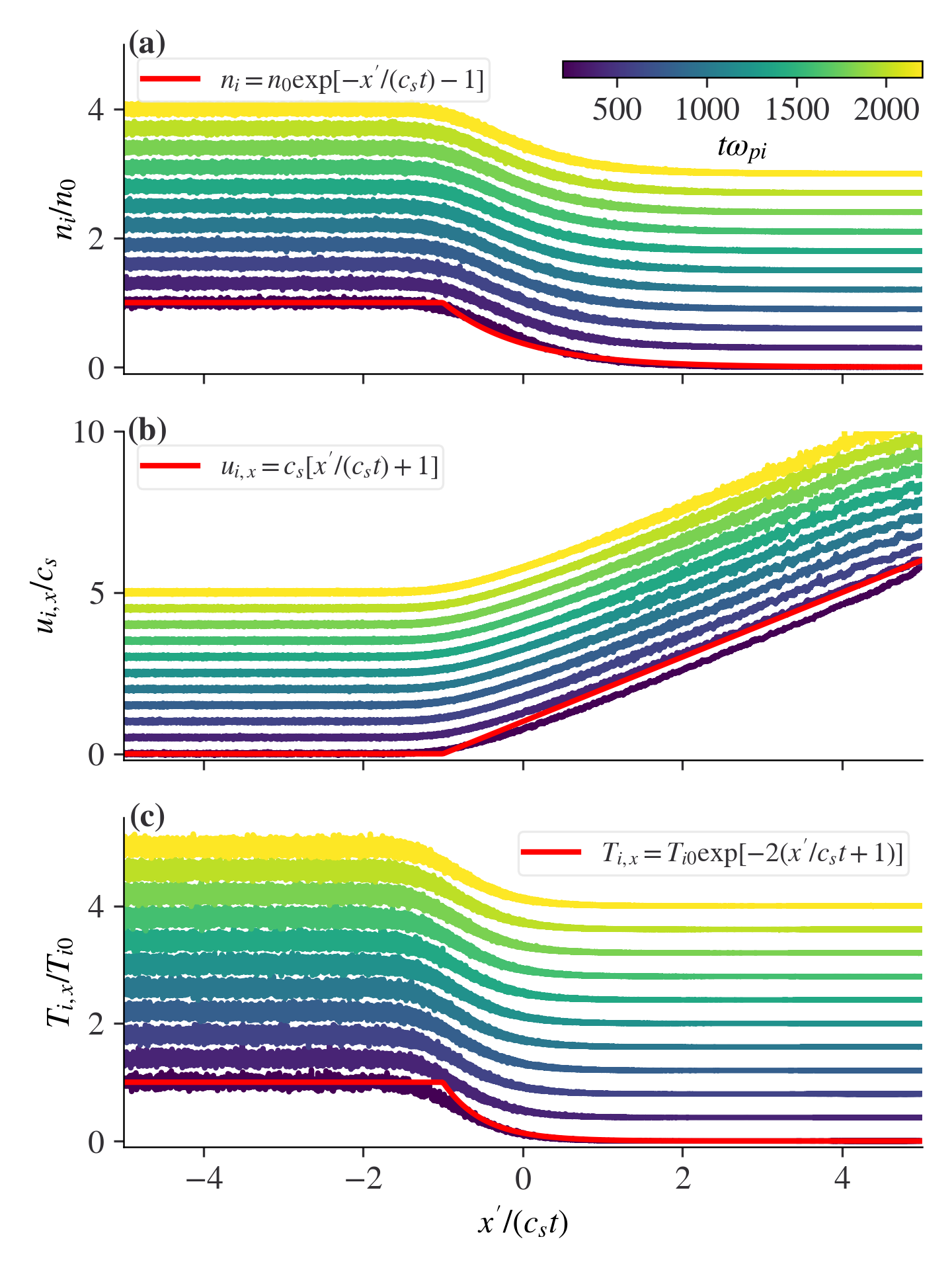}
    \caption{Profiles of plasma density (a), parallel ion flow velocity (b), and parallel ion temperature (c) in the self-similar coordinate $x'/(c_s t)$ at times $t \omega_{pi} = 200, 400, \cdots, 2200$. Here we shift the origin of the coordinate to one edge of the wake as $x' = x + R_l$ for convenience. $x' / (c_s t) = -1$ marks the boundary between unperturbed and expanding plasmas. The profiles at different times are progressively offset to visibility. Red lines represent theoretical predictions.}
    \label{fig:theory-PIC-comparison}
\end{figure}

We compare our simulation results (prior to the encounter of the two plasma streams at the wake's center) with the theoretical predictions in Section \ref{sec:theory}. According to theory, plasma density, ion flow velocity, and ion temperature should follow self-similar solutions. Our simulation confirms this, showing excellent agreement: plasma density decreases exponentially [Figure \ref{fig:theory-PIC-comparison}(a)], parallel ion flow velocity increases linearly [Figure \ref{fig:theory-PIC-comparison}(b)], and parallel ion temperature drops exponentially [Figure \ref{fig:theory-PIC-comparison}(c)]. Furthermore, the spatial profiles of these quantities retain their characteristic shapes when expressed in the self-similar coordinate $x'/(c_s t)$ across different time steps, validating the self-similar nature of plasma expansion into vacuum. We attribute the minor discrepancies between simulation results and self-similar solutions (e.g., the offset between simulated ion flow velocity and theoretical prediction in the expansion region in Figure \ref{fig:theory-PIC-comparison}(b)) to the violation of the isothermal electron assumption in our theoretical model, as demonstrated in Section \ref{subsec:eeos}.

\subsection{Ion front dynamics}
\begin{figure}[tphb]
    \centering
    \includegraphics[width=0.75\linewidth]{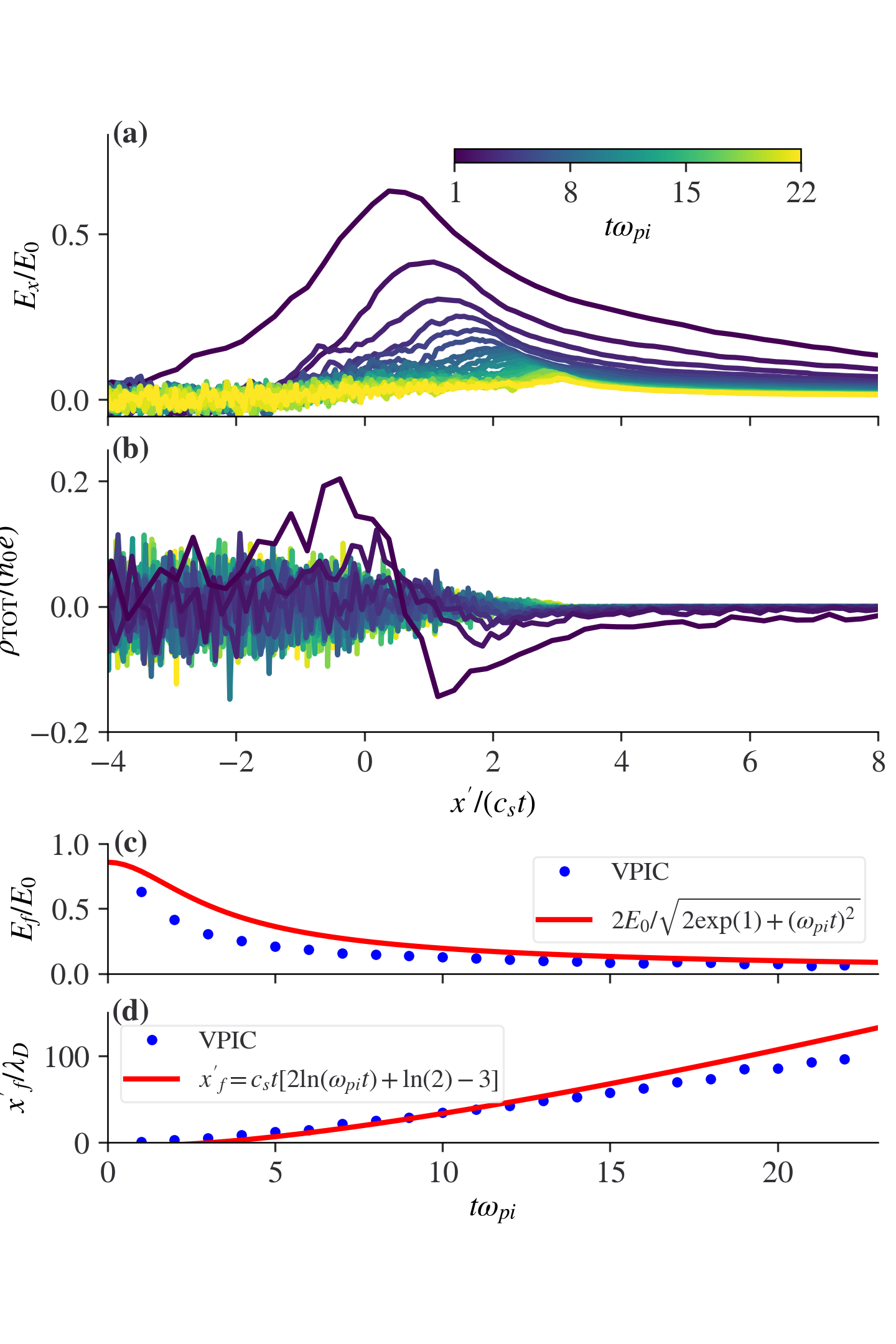}
    \caption{Electric field at the ion front. (a) Electric field profiles\add{ and (b) corresponding total charge density profiles} in the coordinate $x'/(c_s t)$ at times $\omega_{pi} t = 1, 2, \cdots, 22$. The ion front coincides with the peak electric field\add{, as well as the transition from $\rho_{\mathrm{TOT}} > 0$ to $\rho_{\mathrm{TOT}} < 0$}. \change{(b)}{(c)} Comparison of electric field at the ion front from VPIC simulation and theoretical prediction by \citeA{mora2003plasma}. \change{(c)}{(d)} Comparison of ion front location from VPIC simulation and theoretical prediction by \citeA{mora2003plasma}.}
    \label{fig:theory-PIC-Efield}
\end{figure}

The ion front marks the location where the local Debye length exceeds the density gradient scale length, causing the quasi-neutrality condition to break down. Theory predicts the ion front location $x_f = c_s t [2\ln (\omega_{pi} t) - 1]$ [Equation \eqref{eq:ss-xfront}] with an associated electric field peak $E_f = 2 E_0 / (\omega_{pi} t)$ [Equation \eqref{eq:efld-ifront}], where $E_0 = T_{e} / (e \lambda_{D0})$. Figure \ref{fig:theory-PIC-Efield} demonstrates that our simulation captures this electric field propagating with the ion front. The \change{peak}{maximum} electric field \remove{amplitude} follows the predicted $1/(\omega_{pi} t)$ scaling in the large $\omega_{pi} t$ regime.\add{ The breakdown of the quasi-neutrality condition around the ion front is demonstrated in Figure }\ref{fig:theory-PIC-Efield}\add{(b). Across the ion front, the total charge density transitions from positive ($\rho_{\mathrm{TOT}} > 0$) to negative ($\rho_{\mathrm{TOT}} < 0$), consistent with the location of maximum electric field right at the ion front ($\partial E_x / \partial x = 4 \pi \rho_{\mathrm{TOT}} = 0$).} The ion front location, determined from the \change{peak}{maximum} electric field position, approximately follows the\remove{ modified} theoretical prediction $x_f = c_s t [2 \ln(\omega_{pi} t) + \ln 2 - 3]$\add{ by }\citeA{mora2003plasma}.

\remove{Notably, we adopt the ion front location and associated electric field formulations from}\citeA{mora2003plasma}\add{'s prediction $x_f = c_s t [2 \ln(\omega_{pi} t) + \ln 2 - 3]$}\change{, who directly solved Poisson's equation instead of assuming quasi-neutrality, yielding better agreement with our simulation.}{ based on the Poisson's equation is more accurate than the self-similar solution $x_f = c_s t [2\ln (\omega_{pi} t) - 1]$ derived from the approximate quasi-neutrality condition $n_i = n_e$ (Section 2), since quasi-neutrality breaks down around the ion front. However, even Mora's solution overestimates both the maximum electric field and the ion front location [Figures }\ref{fig:theory-PIC-Efield}\add{(c) and }\ref{fig:theory-PIC-Efield}\add{(d)]. This overestimation occurs because electron cooling reduces the electric field at the ion front and consequently slows ion front acceleration, whereas }\citeA{mora2003plasma}\add{ assumed isothermal electrons, as further discussed in Section }\ref{subsec:eeos}\add{.}

\subsection{Electron equation of state}\label{subsec:eeos}
\begin{figure}[tphb]
    \centering
    \includegraphics[width=0.85\linewidth]{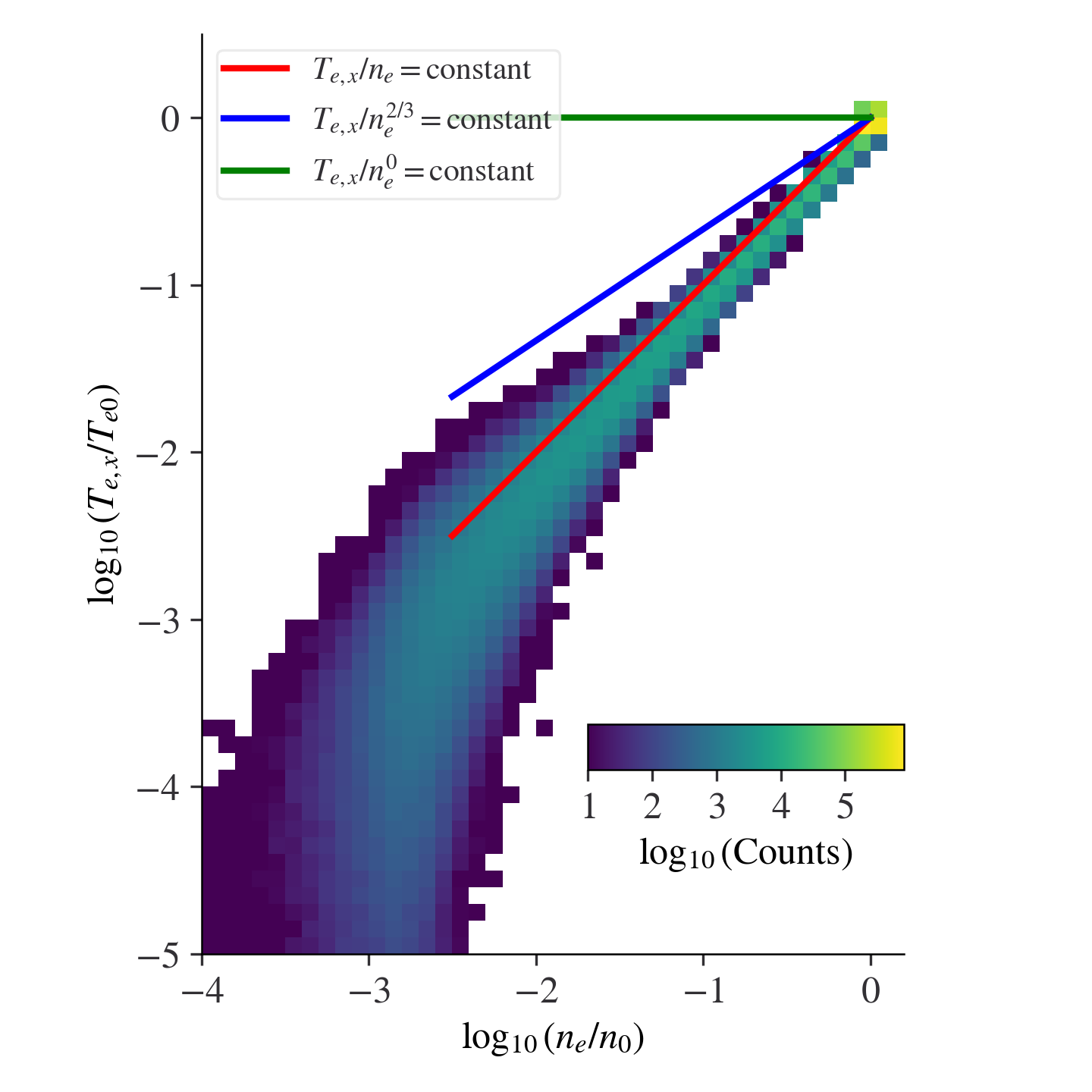}
    \caption{Distribution of electron macroscopic states in the electron density-temperature space. Data are collected from the entire simulation domain at $t \omega_{pi} = 200, 400, \cdots, 2200$. The blue and green lines represent the adiabatic $T_{e,x} / n_e^{2/3} = \mathrm{constant}$ and isothermal $T_{e,x} = \mathrm{constant}$ equations of state, respectively, while the red line corresponds to $T_{e,x} / n_e = \mathrm{constant}$. The ``hot spot'' at $T_{e,x}=T_{e0}$ and $n_e=n_0$ denotes the unperturbed plasma.}
    \label{fig:EOS-electron}
\end{figure}

\begin{figure}
    \centering
    \includegraphics[width=0.8\linewidth]{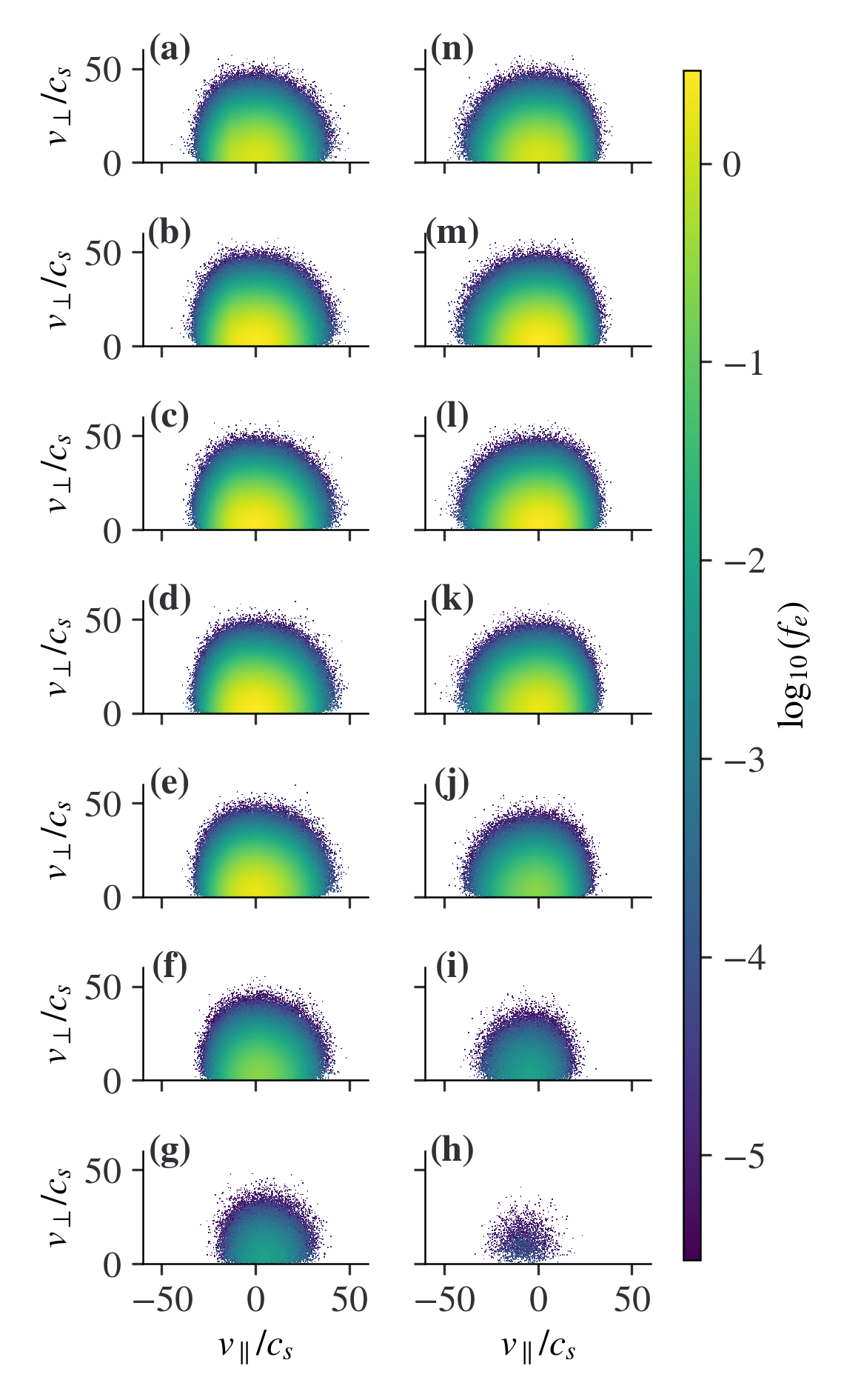}
    \caption{\add{Electron elocity distributions in the $(v_\parallel, v_\perp)$ plane at $t \omega_{pi} = 2200$. Electrons at a certain spatial range are deposited to obtain a velocity distribution in each panel. Panels (a), (b), $\cdots$, and (g) correspond to $-30 \leq x/d_i < -26$, $-26 \leq x/d_i < -22$, $\cdots$, and $-6 \leq x/d_i < -2$, respectively. Panels (h), (i), $\cdots$, and (n) correspond to $2 \leq x/d_i < 6$, $6 \leq x/d_i < 10$, $\cdots$, and $26 \leq x/d_i < 30$, respectively. Panels progress from top to bottom moving spatially from the solar wind into the wake's center. The logarithm of electron phase space density is colored-coded. The observed asymmetry in electron velocity distribution functions across the wake center is likely caused by imperfect symmetry in the initial electron distributions in both configuration and velocity space.}}
    \label{fig:electron-vcl}
\end{figure}

A key assumption in plasma expansion theory \cite<e.g.,>[]{gurevich1966self,mora2003plasma} is that electrons remain isothermal, with $T_e = \mathrm{constant}$. To assess the validity of this assumption, we analyzed electron density and temperature data at the times shown in Figure \ref{fig:theory-PIC-comparison}, with their probability distribution presented in Figure \ref{fig:EOS-electron}. The results show that, similar to electron density, electron temperature decreases exponentially in the expansion direction.\add{ This cooling is also evident in the electron velocity distributions of Figure }\ref{fig:electron-vcl}\add{.} The electron equation of state close to the wake's edge can be approximated as $T_{e, x} / n_e = \mathrm{constant}$ or $P_{e, x} / n_e^2 = \mathrm{constant}$ ($P_{e, x}$ being the electron pressure in the expansion direction), differing from both the isothermal ($P_{e, x}/n_e = \mathrm{constant}$) and adiabatic ($P_{e, x}/n_e^{5/3} = \mathrm{constant}$) cases.\add{ The cooling of electrons in the velocity component parallel to the background magnetic field is also evident in velocity distribution functions shown in Figure }\ref{fig:electron-vcl}\add{. The equation of state indicates that electrons cool more rapidly than would be expected from adiabatic expansion, with $T_{e,x} \propto n_e$ rather than $T_{e, x} \propto n_e^{2/3}$. This behavior is governed by the interplay between pressure-driven expansion and energy loss through work done against the ambipolar electric field. As the plasma expands into vacuum, the ambipolar field that maintains quasi-neutrality leads to electron cooling beyond the adiabatic limit.}

Moreover, further from the wake's edge, as electron properties begin to deviate significantly from their unperturbed values ($n_e / n_0 < 10^{-2}$), our simulation indicates that electrons start to cool more rapidly than the density decay rate, with $T_{e,x}/T_{e0} < n_e / n_0$. Remarkably, despite this very significant deviation from the isothermal electron assumption, theoretical predictions still capture the essential expansion dynamics reasonably well [Figures \ref{fig:theory-PIC-comparison} and \ref{fig:theory-PIC-Efield}]. However, discrepancies between theory and simulation--such as deviations in ion \change{moments near the rarefaction wave front}{flow velocity} [Figure \ref{fig:theory-PIC-comparison}\add{(b)}] and differences in the electric field near the ion front\add{ and the ion front location} [Figure\add{s} \ref{fig:theory-PIC-Efield}\change{(b)}{(c) and }\ref{fig:theory-PIC-Efield}\add{(d)}]--likely originate from this non-isothermal electron behavior.\add{ Specifically, the electron cooling described by $T_{e,x} / n_e = \mathrm{constant}$ reduces the electron pressure gradient that generates the ambipolar electric field driving ion acceleration, leading to weaker electric fields, lower ion velocities, and smaller ion front locations than predicted by the isothermal model.}

\section{Electrostatic shocks generated by counter-streaming ion beams}\label{sec:electrostatic-shocks}
\subsection{Propagation characteristics}
\begin{figure}[tphb]
    \centering
    \includegraphics[width=\linewidth]{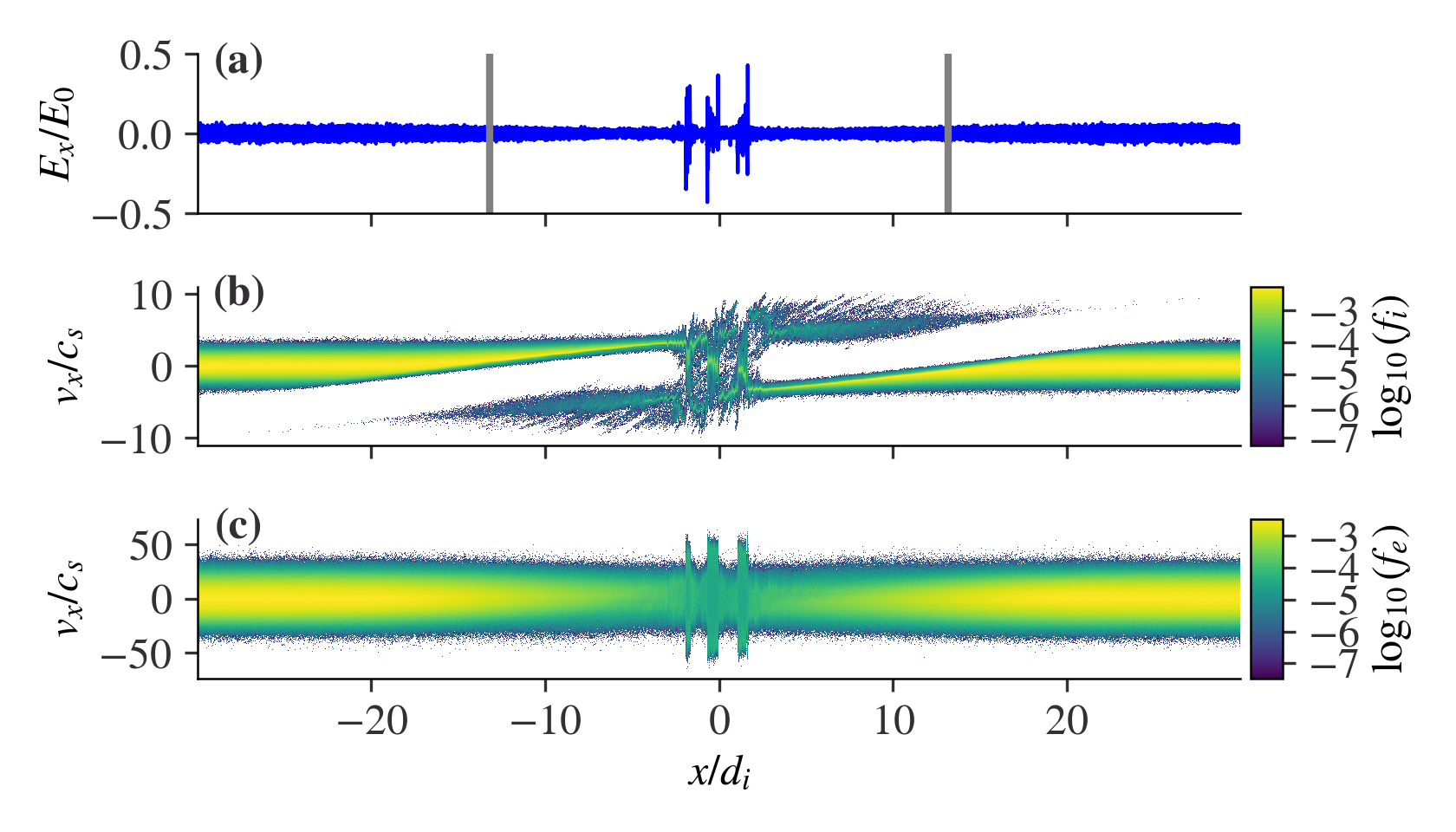}
    \caption{Electric field and particle phase space densities at time $t \omega_{pi} = 7500$. (a) Electric field normalized to $E_0 = T_e / (e \lambda_{D0})$. The two gray lines indicate positions at one lunar radius from the wake center, $x = \pm R_l$. (b) Ion distribution in the phase space $(x, v_x)$ (c) Electron distributions in the phase space $(x, v_x)$.}
    \label{fig:ps-snapshot}
\end{figure}

\begin{figure}[tphb]
    \centering
    \includegraphics[width=0.85\linewidth]{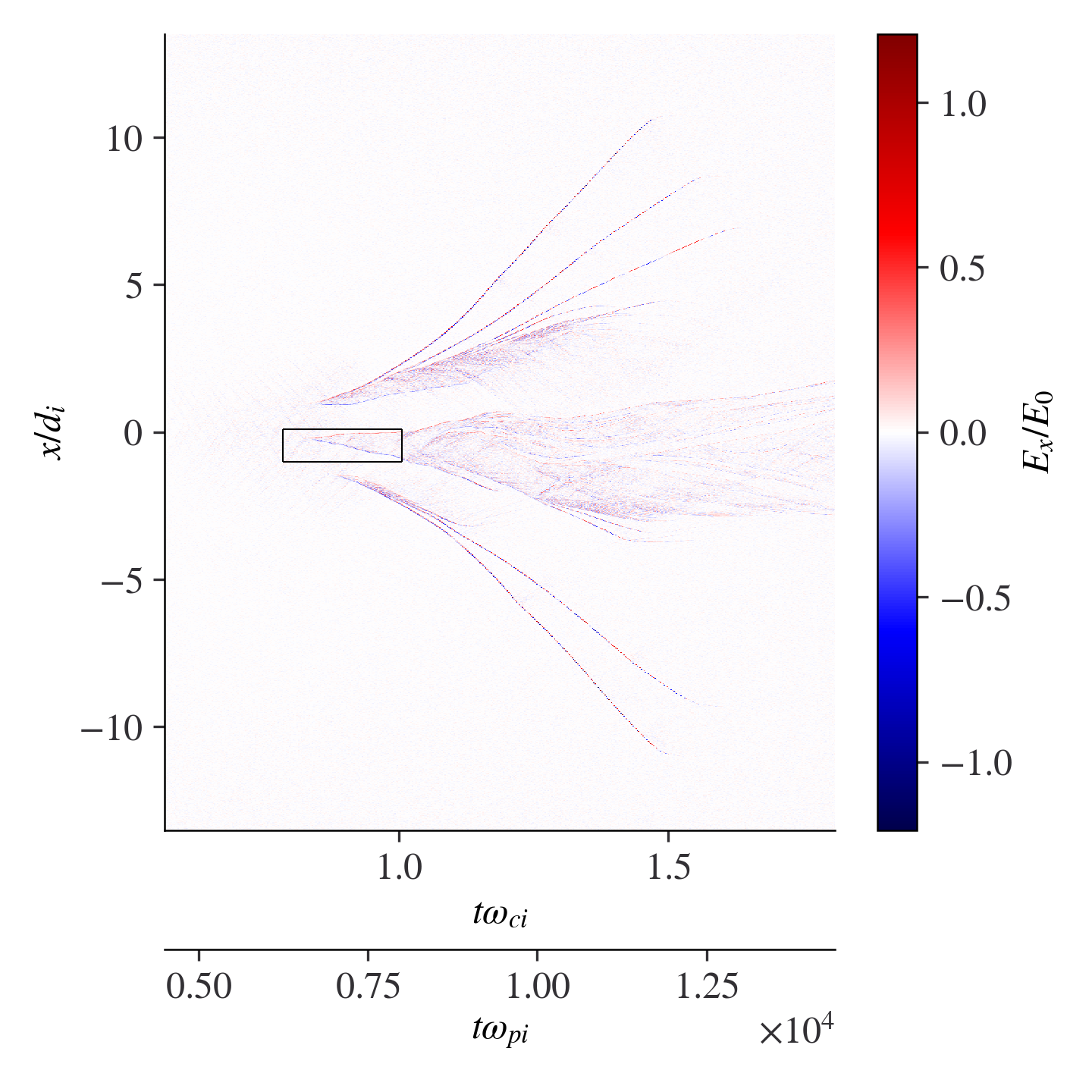}
    \caption{Spatiotemporal evolution of the electric field. The spatial domain, $-13.5 \leq x / d_i \leq 13.5$, approximately corresponds to the lunar diameter and is centered on the wake center. The black box highlights the specific region analyzed in detail in Figure \ref{fig:wave-xt-center}(a).}
    \label{fig:tjoin-efield}
\end{figure}

At about $t \omega_{pi} = 7500$, the two counter-streaming, supersonic ion beams expanding inwards from the two sides of the wake ($u_+ = -u_- = 5 c_s = 25 c_{s, \mathrm{local}}$) with appreciable density ($n_+ = n_- \simeq 0.025$) collide at the wake center [Figures \ref{fig:ps-snapshot}(b) and \ref{fig:tjoin-density}]. This collision generates three groups of electrostatic shocks: those propagating in the $+x$ direction, those propagating in the $-x$ direction, and those propagating bidirectionally in both the $+x$ and $-x$ directions, as shown in Figure \ref{fig:tjoin-efield}. The electric field associated with these shocks can reach nearly $0.5 E_0 = 0.5 T_e / (e \lambda_{D0})$ [Figure \ref{fig:ps-snapshot}(a)], where $T_e$ is the electron temperature and $\lambda_{D0}$ is the Debye length in the solar wind. Downstream of these shocks, the ion beams decelerate [Figure \ref{fig:ps-snapshot}(b)] while electrons undergo significant heating [Figure \ref{fig:ps-snapshot}(c)].

\begin{figure}[tphb]
    \centering
    \includegraphics[width=\linewidth]{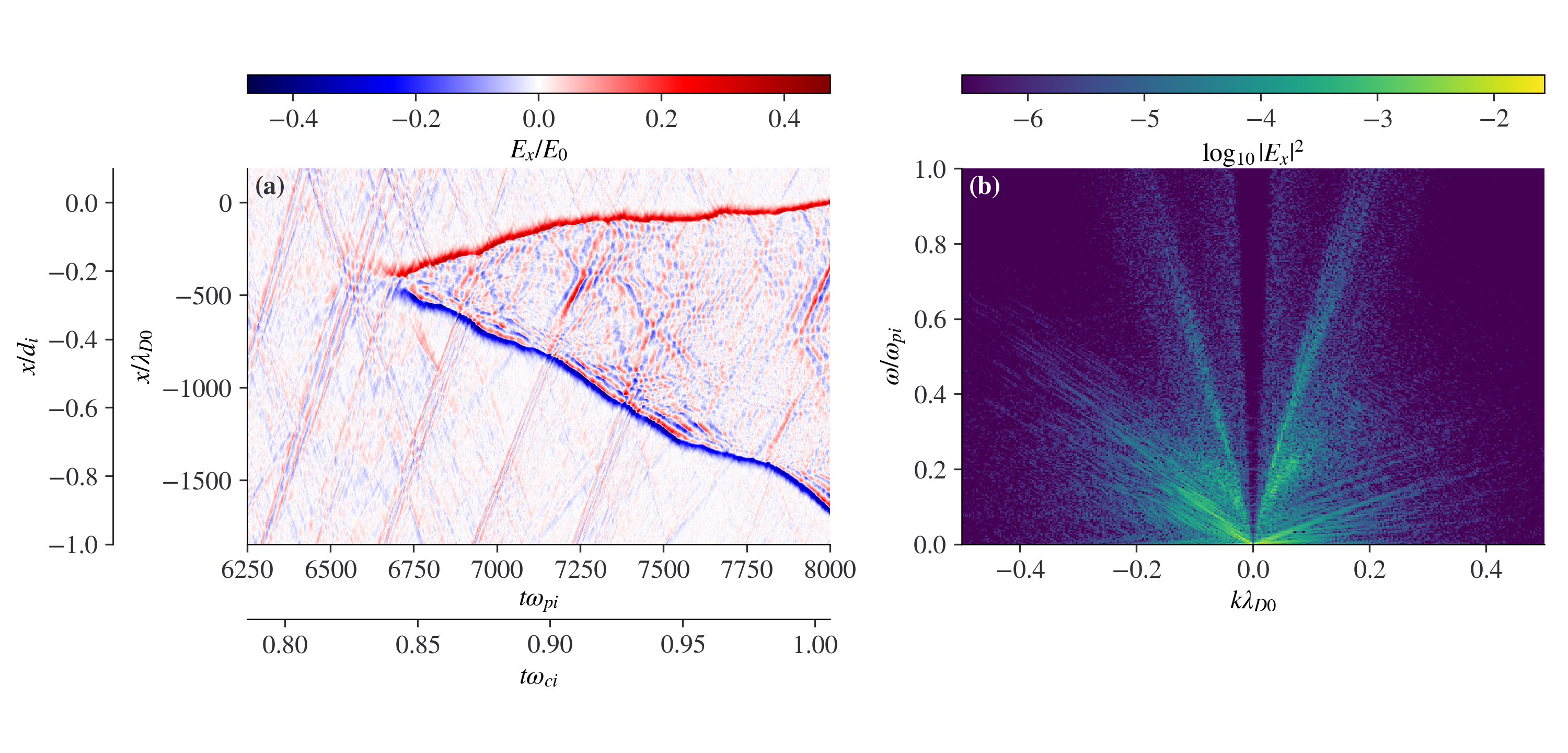}
    \caption{Propagation characteristics of electrostatic shocks and beam-mode ion acoustic waves. (a) Spatiotemporal evolution of bidirectionally propagating shocks in Figure \ref{fig:tjoin-efield}. (b) Electric field power density in wavenumber-frequency $(k, \omega)$ space.}
    \label{fig:wave-xt-center}
\end{figure}

The formation mechanisms of the three shock groups seen in Figures \ref{fig:ps-snapshot}(a) and \ref{fig:tjoin-efield} \change{are similar}{can be attributed to the ion acoustic instability driven by supersonic ion beams, regardless of whether they are symmetric or asymmetric with respect to $v_x=0$. The observed asymmetric shock structures with respect to $x=0$ in Figure }\ref{fig:tjoin-efield}\add{ likely result from asymmetries in the ion beam distributions on either side of the wake, which can arise from small deviations from perfect symmetry in the initial particle distributions. The electron distribution functions similarly exhibit this imperfect symmetry [Figure }\ref{fig:electron-vcl}\add{]}. Using the bidirectionally propagating shock as an example, shown in detail in Figure \ref{fig:wave-xt-center}(a), we analyze its characteristics. By decomposing the electric field in the spatiotemporal domain into Fourier modes, $E_x (x, t) = \iint \mathrm{d}\omega\, \mathrm{d}k\, E_x(k, \omega) \exp\left(i k x - i \omega t\right)$, we show the power density $\vert E_x(k, \omega) \vert^2$ in Figure \ref{fig:wave-xt-center}(b). The electric field energy is predominantly concentrated within the range $-0.2 < k \lambda_{D0} < 0.2$ (equivalent to $-0.9 < k \lambda_{D,\mathrm{local}} < 0.9$ due to the enlarged local Debye length $\lambda_{D,\mathrm{local}} = 4.5 \lambda_{D0}$) and $0 < \omega / \omega_{pi} < 0.4$. These wavenumber and frequency ranges are characteristic of ion-acoustic modes \cite<e.g.,>[]{davidson1970electron,gary1987ion}.

\begin{figure}[tphb]
    \centering
    \includegraphics[width=\linewidth]{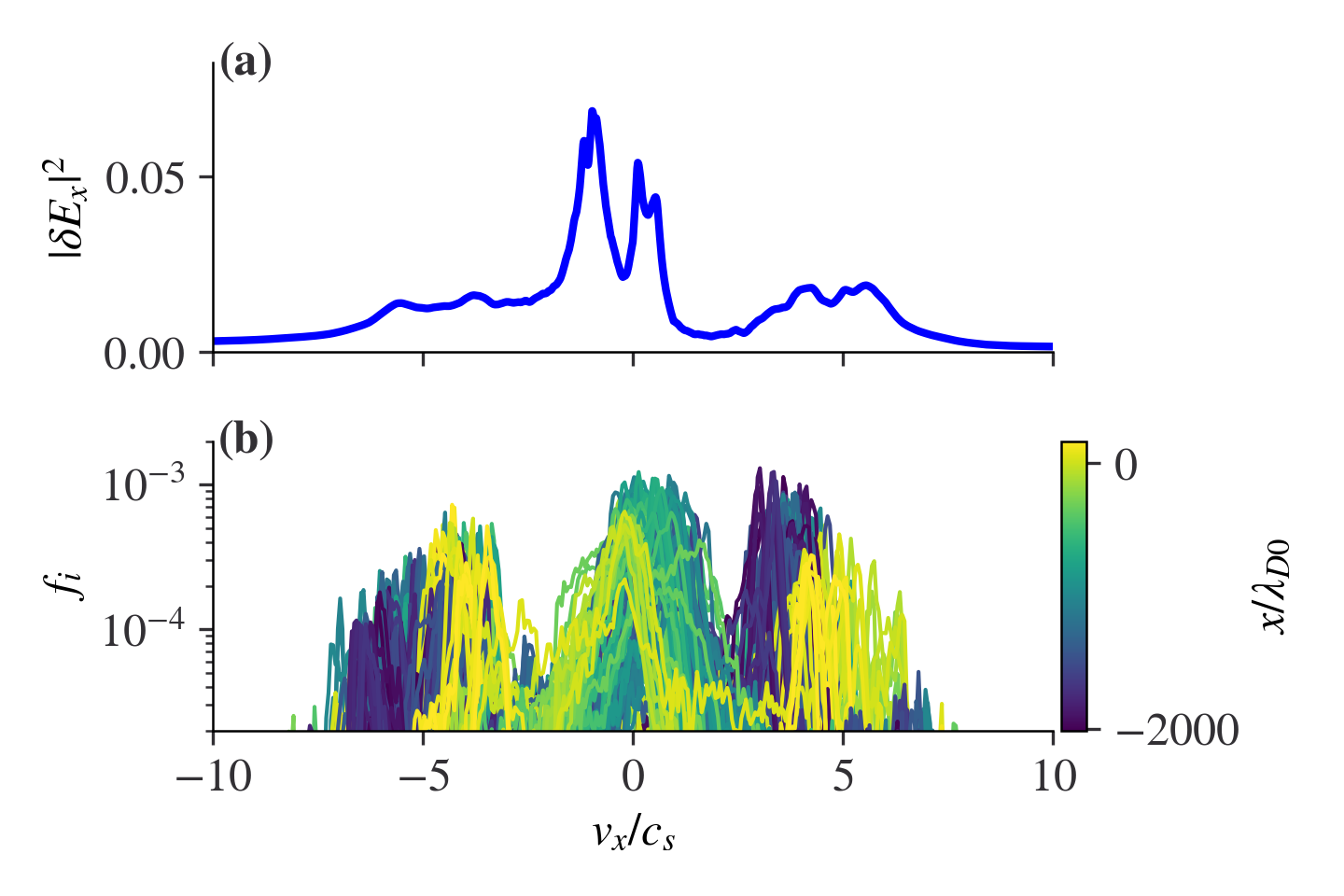}
    \caption{Phase velocities of the excited electric field structures and their associated ion velocity distributions. (a) Electric field power density as a function of phase velocities. (b) Ion velocity distribution functions measured at spatial positions corresponding to Figure \ref{fig:wave-xt-center} at selected times $t \omega_{pi} = 6500, 6750, 7000, 7250, 7500, 7750, 8000$.}
    \label{fig:wave-ps-center}
\end{figure}

The electric field energy is distinctly organized around four well-separated phase velocities in $(k, \omega)$ space, as shown Figure \ref{fig:wave-xt-center}(b). By integrating the electric field energy for each phase velocity $\omega / k$, we obtain $\vert E_x \vert^2$ as a function of $\omega / k$ [Figure \ref{fig:wave-ps-center}(a)]. The dominant energy contribution comes from two shocks propagating at $\omega / k c_s = -1$ and $\omega / k c_s = 0.1$--$0.5$. Secondary energy peaks appear around $\omega / k c_s = \pm 5$. These faster-propagating waves, existing prior to shock formation [Figure \ref{fig:wave-xt-center}(a)], are ion-acoustic modes driven by ion beams centered at $v_x / c_s = \pm 5$ [Figure \ref{fig:wave-ps-center}(b)]. The shocks themselves correspond to a slower ion population concentrated in the range $-1 < v_x/c_s < 1$ [Figure \ref{fig:wave-ps-center}(b)], which is formed when the shocks decelerate the incoming beams moving at $\sim \pm 5 c_s$, as demonstrated below.

\subsection{Energy conversion}\label{subsec:energy-conversion}
\begin{figure}[tphb]
    \centering
    \includegraphics[width=\linewidth]{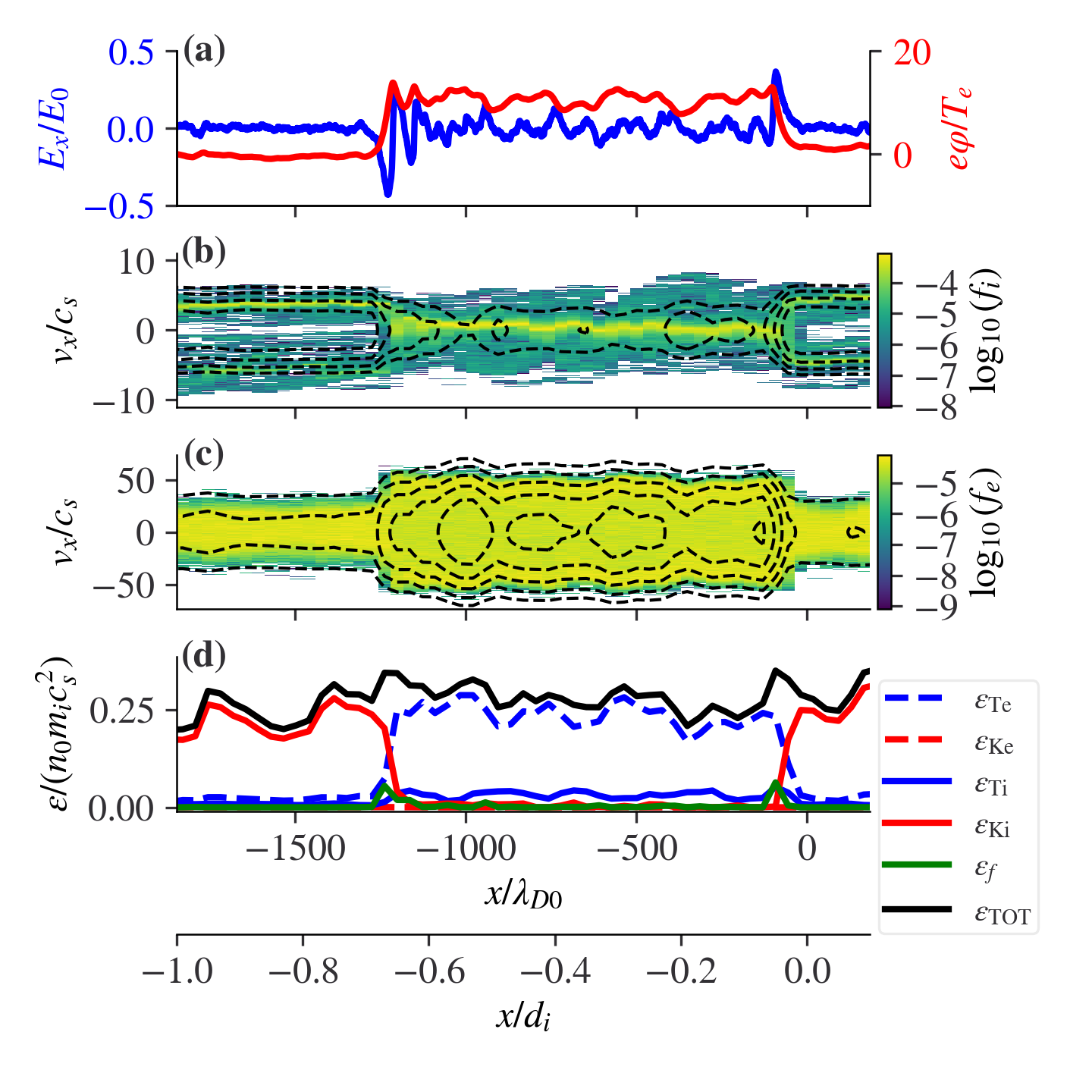}
    \caption{Energy conversion across electrostatic shocks\add{ at $t \omega_{pi} = 7500$}. (a) Spatial profiles of the electric field $E_x$ (blue) and the associated potential $\varphi = -\int E_x \mathrm{d}x$ (red). (b) Ion phase-space distribution in the position-velocity $(x, v_x)$ plane. Dashed lines represent contours of the ion Hamiltonian $H_i$ from Equation \eqref{eq:Hi}. (c) Electron phase-space distribution in the $(x, v_x)$ plane. Dashed lines represent contours of the electron Hamiltonian $H_e$ from Equation \eqref{eq:He}. (d) Partitioning of energy among different forms (kinetic, thermal, field) across the shock structure. Note: In the upstream regions\add{ ($x/d_i < -0.64$ or $x/d_i \geq -0.04$)}, we separate the incoming and outgoing ion beams, i.e., the $v_x > 0$ and $v_x < 0$ populations, to calculate their respective kinetic and thermal energy densities. This separation prevents the relative drift between counter-streaming beams from being misinterpreted as thermal energy.}
    \label{fig:ps-zoom-ene}
\end{figure}

To understand the energy conversion across the shocks, we examine their structures at $t \omega_{pi} = 7500$, as shown in Figure \ref{fig:ps-zoom-ene}(a). Across each of the shock surface, over a short distance of $\sim 90 \lambda_{D0}$ (equivalent to $20 \lambda_{D, \mathrm{local}}$) from upstream to downstream, the electric potential increases by $\Delta \varphi \sim 10 T_e / e$, creating an electric field directed toward the upstream. This electric field either decelerates the incoming ion beams to subsonic speeds in the downstream region or reflects ions back upstream if they lack sufficient energy to overcome the potential barrier [Figure \ref{fig:ps-zoom-ene}(b)]. The ion motion follows the Hamiltonian
\begin{linenomath*}
    \begin{equation}\label{eq:Hi}
        H_i = \frac{1}{2} m_i v_{i,x}^2 + e \varphi ,
    \end{equation}
\end{linenomath*}
which can be used to estimate the potential difference across the shock. Since the bulk ions transition from highly supersonic upstream to nearly zero downstream in our system [Figure \ref{fig:ps-zoom-ene}(b)], $v_{i,\mathrm{up}}^2 \gg v_{i,\mathrm{down}}^2$ (where $v_{i, \mathrm{up}}$ and $v_{i, \mathrm{down}}$ represent ion velocities upstream and downstream, respectively), yielding
\begin{linenomath*}
    \begin{equation}
        \Delta \varphi = \frac{m_i v_{i, \mathrm{up}}^2}{2 e} ,
    \end{equation}
\end{linenomath*}
or in normalized form
\begin{linenomath*}
    \begin{equation}\label{eq:shock-potential}
        \frac{e \Delta \varphi}{T_e} = \frac{1}{2} \left(\frac{v_{i, \mathrm{up}}}{c_s}\right)^2 .
    \end{equation}
\end{linenomath*}
For $v_{i, \mathrm{up}} \sim \pm 5 c_s$, this equation gives $e \Delta \varphi / T_e \sim 12.5$, consistent with our simulation results [Figure \ref{fig:ps-zoom-ene}(a)].

Electrons are accelerated through the shock potential [Figure \ref{fig:ps-zoom-ene}(c)], with their trajectories governed by the Hamiltonian
\begin{linenomath*}
    \begin{equation}\label{eq:He}
        H_e = \frac{1}{2} m_e v_{e,x}^2 - e \varphi .
    \end{equation}
\end{linenomath*}
The shock potential $\Delta \varphi$ creates a hole in the velocity distribution, which rapidly fills through nonlinear trapping to form a flat-top distribution. The half-width velocity of this plateau is given by
\begin{linenomath*}
    \begin{equation}
        \Delta v_e = \sqrt{\frac{2 e \Delta \varphi}{m_e}} ,
    \end{equation}
\end{linenomath*}
or in normalized form
\begin{linenomath*}
    \begin{equation}\label{eq:width-flattop}
        \frac{\Delta v_e}{c_s} = \sqrt{\frac{m_i}{m_e} \cdot \frac{2 e \Delta \varphi}{T_e}} .
    \end{equation}
\end{linenomath*}
With $m_i / m_e = 100$ and $e \Delta \varphi / T_e \sim 12.5$, we calculate $\Delta v_e / c_s \sim 50$, which captures the half width of the flat-top distribution observed downstream [Figure \ref{fig:ps-zoom-ene}(c)].

Energy conservation in our system can be expressed as \cite{davidson2012methods}
\begin{linenomath*}
    \begin{equation}
        \int \mathrm{d}x\,\varepsilon_{\mathrm{TOT}} = \int \mathrm{d}x\, \left( \sum_{s = i, e} \underbrace{\frac{1}{2} n_s m_s u_{x,s}^2}_{\varepsilon_{Ks}} + \sum_{s = i, e} \underbrace{\frac{1}{2} n_s m_s v_{Tx,s}^2}_{\varepsilon_{Ts}} + \underbrace{\frac{E_x^2}{8 \pi}}_{\varepsilon_f} \right) = \mathrm{constant} ,
    \end{equation}
\end{linenomath*}
where $n_s$ is the density of species $s$, $\varepsilon_{Ks}$ represents the kinetic energy density of species $s$, $\varepsilon_{Ts}$ denotes the thermal energy density of species $s$, $\varepsilon_f$ is the electric field energy density, and $\varepsilon_{\mathrm{TOT}}$ is the total energy density. Figure \ref{fig:ps-zoom-ene}(d) illustrates the evolution of these energy density components across the shocks. The data clearly shows that cross shock interfaces, ion kinetic energy is predominantly converted into electron thermal energy, with only a smaller fraction transformed into ion thermal energy. Strong, localized electric fields at the shock surfaces mediate this energy transfer by decelerating bulk ion flows and heating electrons that encounter or traverse these fields. The energy stored in these electric fields is negligible compared to both the upstream ion kinetic energy and the downstream electron thermal energy.

\subsection{Electrostatic shocks at late times}
\begin{figure}[tphb]
    \centering
    \includegraphics[width=\linewidth]{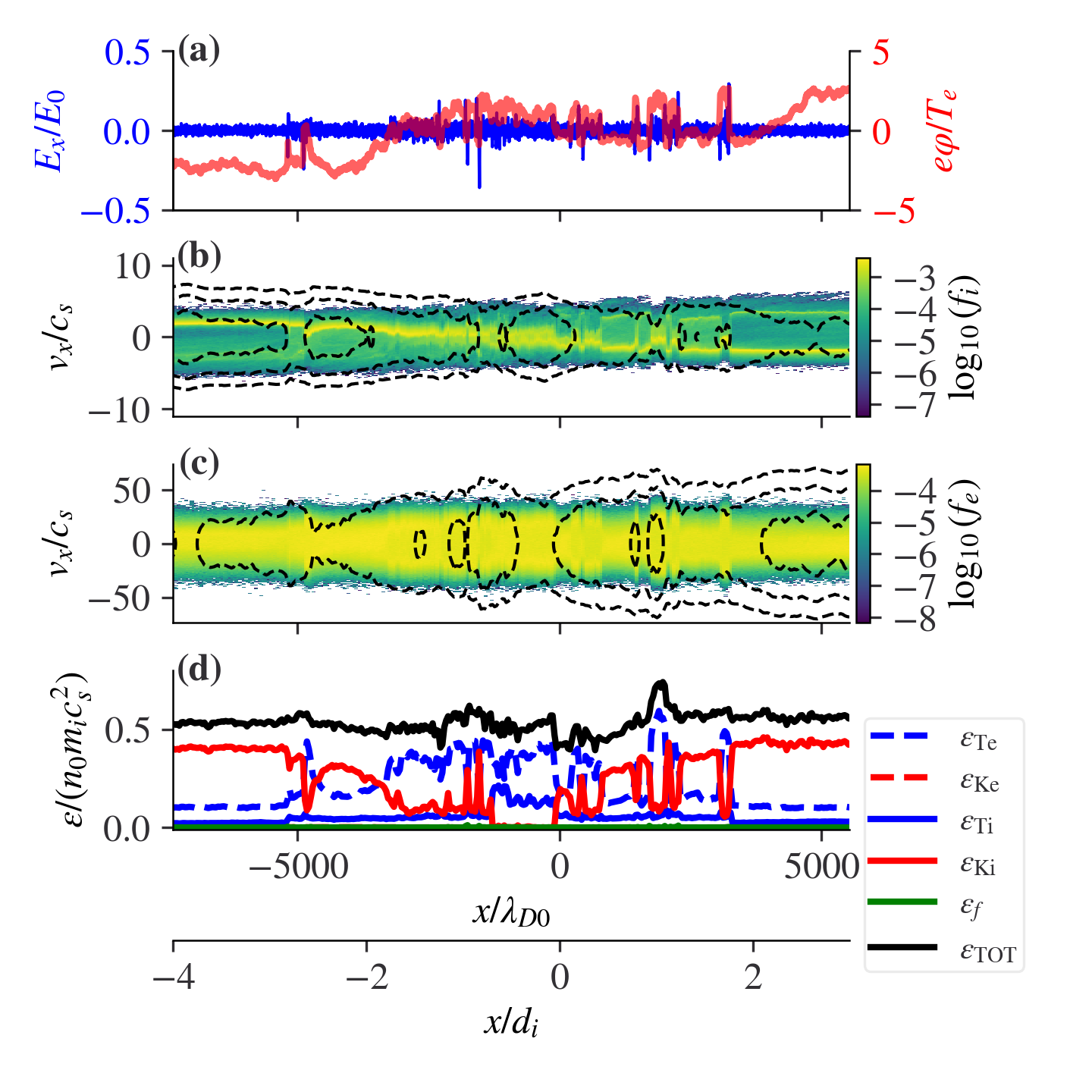}
    \caption{Electric fields, phase space density distributions, and energy partition between fields and particles at $t \omega_{pi} = 14400$. The format is the same as Figure \ref{fig:ps-zoom-ene}.\add{ Note: Since the ion velocity distributions are thermalized, we treat ions at each spatial location as a single population.}}
    \label{fig:ps-snapshot-final}
\end{figure}

Figure \ref{fig:ps-snapshot-final} displays the electric potential and energy partition between fields and particles at the final simulation time ($t \omega_{pi} = 14400$), which is representative of the conditions at later times in our simulation. The spatial region depicted corresponds to the domain evolving from the shock waves presented in Figure \ref{fig:ps-zoom-ene} [see the shock wave propagation in Figure \ref{fig:tjoin-efield}]. Compared to Figure \ref{fig:ps-zoom-ene}(a), the shock potentials in Figure \ref{fig:ps-snapshot-final}(a) exhibit decreased amplitudes and increased structural intermittency. The inhomogeneous plasma is accompanied by a large-scale electric potential gradient. Particle interactions with this large-scale electric potential gradient and concurrent localized solitary shock waves result in nonlinear scattering and trapping of particles [Figures \ref{fig:ps-snapshot-final}(b) and \ref{fig:ps-snapshot-final}(c)]. Such interactions give rise to particle transport and energy transfer analogous to nonlinear particle interactions with solitary waves in the Earth's inner magnetosphere \cite{karpman1975effects,artemyev2017nonlinear}. We will address the formation and evolution of this large-scale electric potential gradient in a separate publication. Consistent with Figure \ref{fig:ps-zoom-ene}(d), Figure \ref{fig:ps-snapshot-final}(d) demonstrates that in regions of enhanced electric field fluctuations, ion kinetic energy is primarily converted to electron thermal energy, while a small fraction becomes ion thermal energy.

\subsection{Impact of reduced ion-to-electron mass ratio and lunar radius}\label{subsec:scaling}
Understanding how electric potential jumps and electron heating across electrostatic shocks scale with parameters such as ion-to-electron mass ratio and lunar radius is crucial for characterizing these systems under realistic conditions. Electrostatic shocks form naturally at the wake's center when two supersonic ion beams collide. Prior to shock formation, we can determine the density and velocity of these beams using Equations \eqref{eq:ss-den} and \eqref{eq:ss-vel}. The critical condition for shock generation occurs when the normalized momentum density reaches a threshold value $n u / (n_0 c_s) \geq \eta_c$ \cite<e.g.,>[]{davidson1970electron}. We can express this critical condition mathematically as
\begin{linenomath*}
    \begin{align}
        \left(\frac{R_l}{c_s t_c} + 1\right) \cdot \exp\left(-\frac{R_l}{c_s t_c} - 1\right) = \eta_c,
    \end{align}
\end{linenomath*}
where $t_c$ represents the time at which the momentum density first exceeds the critical threshold $\eta_c n_0 c_s$, with $\eta_c$ in the range $0 \leq \eta_c \leq \exp(-1)$. By applying the Lambert $W$ function \cite{corless1996lambert,lehtonen2016lambert}, we can invert this equation to solve for $R_l / (c_s t_c)$:
\begin{linenomath*}
    \begin{align}
        \frac{R_l}{c_s t_c} + 1 = -W_{-1}(-\eta_c).
    \end{align}
\end{linenomath*}
\add{The reduced system size ($R_l / \lambda_D$) does decrease the characteristic time ($t_c \omega_{pi}$) required for ion beams from opposite sides of the lunar wake to converge at the wake center. However, the dimensionless parameter $R_l / (c_s t_c)$ remains invariant under this scaling. This invariance ensures that both the densities and velocities of the converging ion beams are preserved. }This allows us to derive the critical density and ion beam velocity:
\begin{linenomath*}
    \begin{align}
        \frac{n_c}{n_0} &= \exp[W_{-1}(-\eta_c)], \\  
        \frac{u_c}{c_s} &= -W_{-1}(-\eta_c). \label{eq:u-critical}  
    \end{align}
\end{linenomath*}
Combining Equations \eqref{eq:u-critical} and \eqref{eq:shock-potential}, we can express the shock potential as:
\begin{linenomath*}
    \begin{align}
        \frac{e \Delta \varphi}{T_e} = \frac{1}{2} W_{-1}^2(-\eta_c).
    \end{align}
\end{linenomath*}
A significant finding is that the shock potential depends exclusively on $T_e$ and $\eta_c$ (the latter controlled by the intrinsic ion acoustic instability), with no dependence on lunar radius. Our PIC simulation indicates $\eta_c \approx 0.05$ [Figures \ref{fig:tjoin-density} and \ref{fig:ps-snapshot}], resulting in $e \Delta\varphi / T_e \approx 10$, which is consistent with Figure \ref{fig:ps-zoom-ene}(a). Additionally, we can rewrite the half-width of the electron flat-top velocity distribution [Equation \eqref{eq:width-flattop}] as:
\begin{linenomath*}
    \begin{align}
        \frac{\Delta v_e}{c_s} = \sqrt{\frac{m_i}{m_e}} |W_{-1}(-\eta_c)| \approx 4.5 \sqrt{\frac{m_i}{m_e}}.
    \end{align}
\end{linenomath*}
Under realistic conditions with $m_i / m_e = 1836$, this yields $\Delta v_e / c_s \approx 193$.

\section{Electromagnetic instabilities of anisotropic ion beams}\label{sec:electromagnetic-instabilities}
\begin{figure}[tphb]
    \centering
    \includegraphics[width=\linewidth]{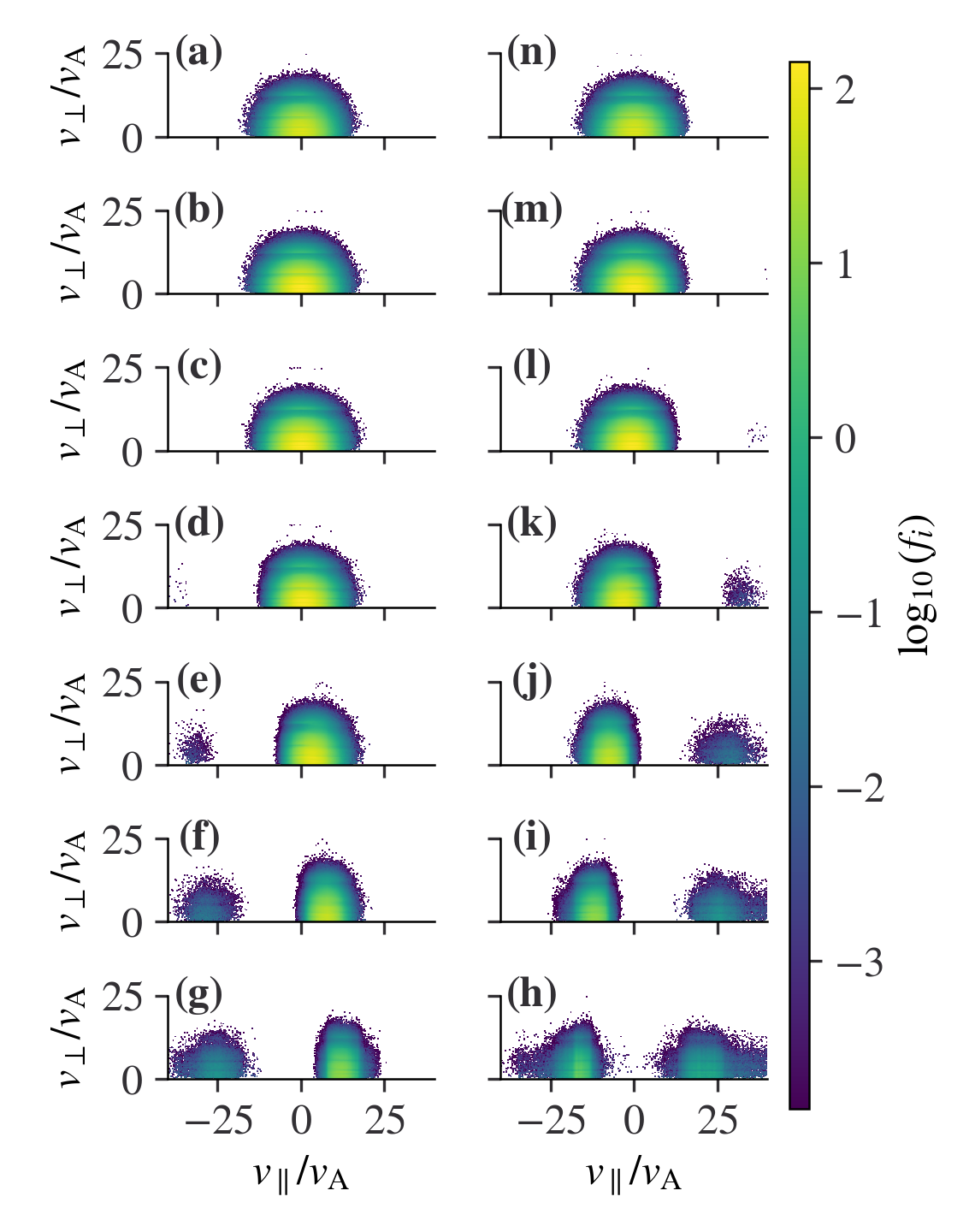}
    \caption{Ion velocity distributions in the $(v_\parallel, v_\perp)$ plane at $t \omega_{pi} = 6000$. Ions at a certain spatial range are deposited to obtain a velocity distribution in each panel. Panels (a), (b), $\cdots$, and (g) correspond to $-30 \leq x/d_i < -26$, $-26 \leq x/d_i < -22$, $\cdots$, and $-6 \leq x/d_i < -2$, respectively. Panels (h), (i), $\cdots$, and (n) correspond to $2 \leq x/d_i < 6$, $6 \leq x/d_i < 10$, $\cdots$, and $26 \leq x/d_i < 30$, respectively. From top to bottom panels, we move from the solar wind to the wake's center. The logarithm of ion phase space density is colored-coded.}
    \label{fig:ion_vcl}
\end{figure}

As ions enter the lunar wake, they form beams along field lines while their parallel temperatures decrease exponentially, creating a perpendicular temperature anisotropy [Equation \eqref{eq:ion-temprature} and Figure \ref{fig:ion_vcl}]. These anisotropic ion beams may trigger electromagnetic instabilities.

\subsection{Magnetosonic waves and anomalous cyclotron resonance}
\begin{figure}[tphb]
    \centering
    \includegraphics[width=\linewidth]{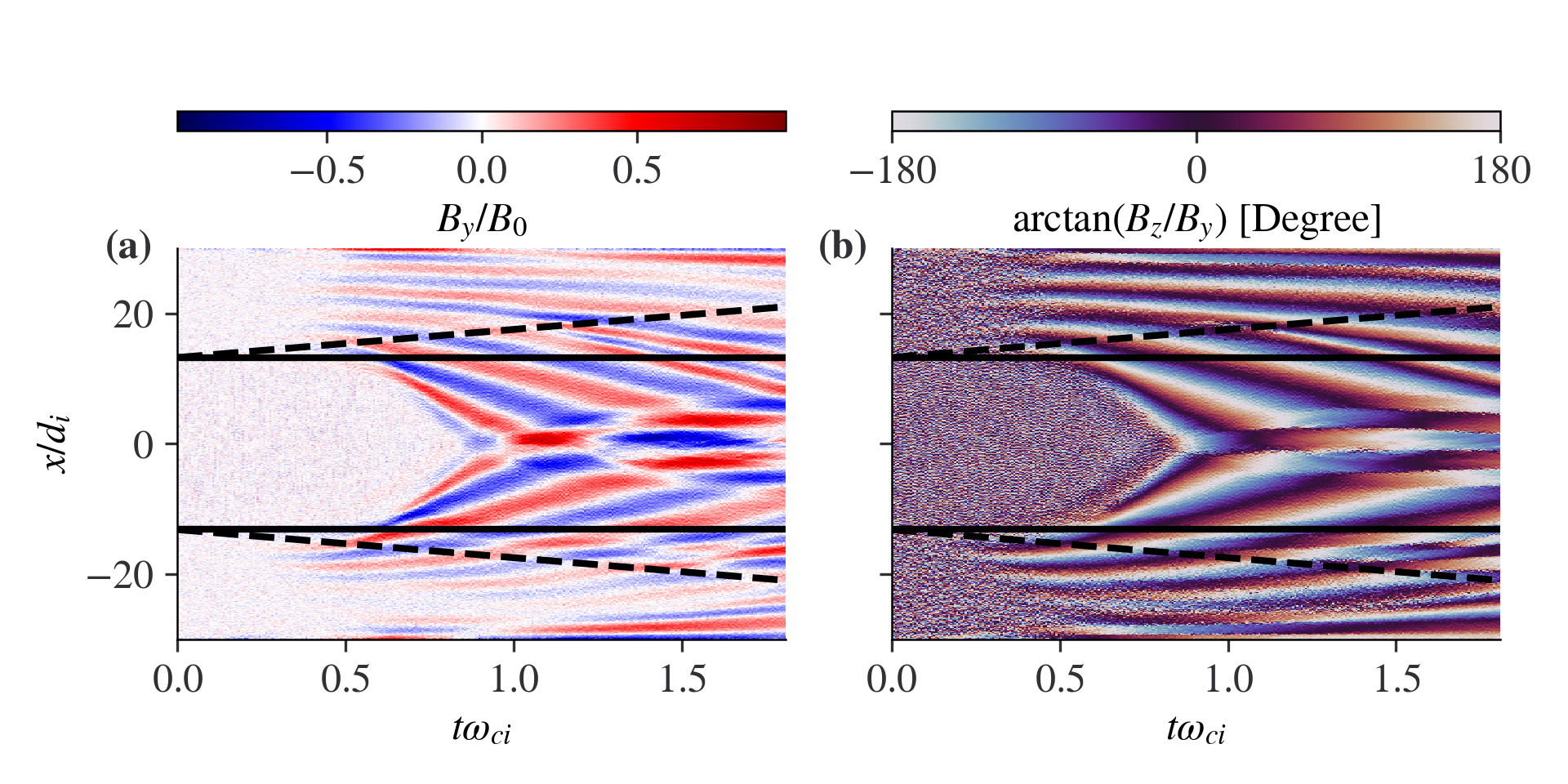}
    \caption{Propagation characteristics of magnetosonic waves. (a) Spatiotemporal evolution of the normalized magnetic field component $B_y / B_0$. (b) Polarization angle $\arctan(B_y/B_z)$ in the $y$-$z$ plane, where $B_y$ leads $B_z$ by $90^{\circ}$ at fixed locations. Note that oscillations near the simulation boundaries arise from boundary effects.\add{ Boundary validation tests with damping layers confirm that the observed ion-scale wave phenomena are physically meaningful.}}
    \label{fig:bwave-polarization}
\end{figure}

\begin{figure}[tphb]
    \centering
    \includegraphics[width=\linewidth]{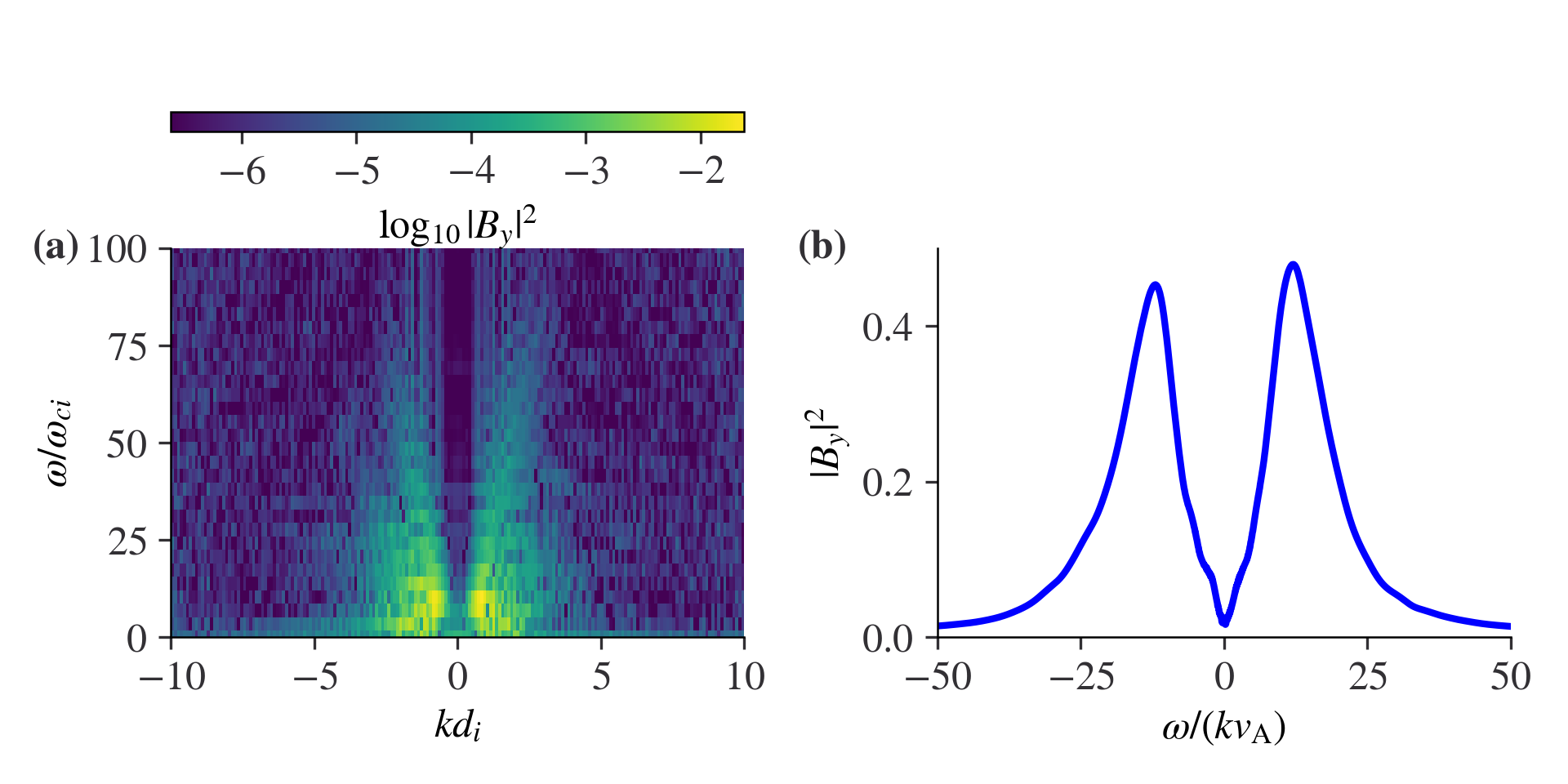}
    \caption{Magnetosonic waves in Fourier space. (a) Magnetic field power spectral density in wavenumber-frequency ($k$, $\omega$) space. (b) Magnetic field power distribution as a function of wave phase velocities.}
    \label{fig:bwave-phasespeed}
\end{figure}

Large-amplitude electromagnetic waves are excited at the wake's edge and propagate inward toward the center [Figure \ref{fig:bwave-polarization}(a)]. During propagation, these waves undergo substantial convective growth, reaching peak amplitudes near the wake's center where waves from opposing sides interfere. Polarization analysis at a fixed location demonstrates that these waves are right-hand polarized with respect to $\mathbf{B}_0$, confirming their identity as magnetosonic waves [Figure \ref{fig:bwave-polarization}(b)]. The magnetosonic waves span frequencies $0< \omega / \omega_{ci} < 40$ and wavenumbers $0 < k d_i < 2.5$ [Figure \ref{fig:bwave-phasespeed}(a)], yielding phase velocities in the range $0 < \vert \omega / (k v_\mathrm{A}) \vert < 25$ [Figure \ref{fig:bwave-phasespeed}(b)].\add{ We use the term ``magnetosonic waves'' to specify the ion-scale ($0 < k d_i < 2.5$) electromagnetic waves relevant to our analysis. These waves belong to the R-mode branch of the dispersion relation and are also referred to as magnetosonic-whistler waves or whistler waves in the literature} \cite{gary1986low,saito2007whistler,gary2008cascade,gary2013test}\add{.}

\begin{figure}[tphb]
    \centering
    \includegraphics[width=\linewidth]{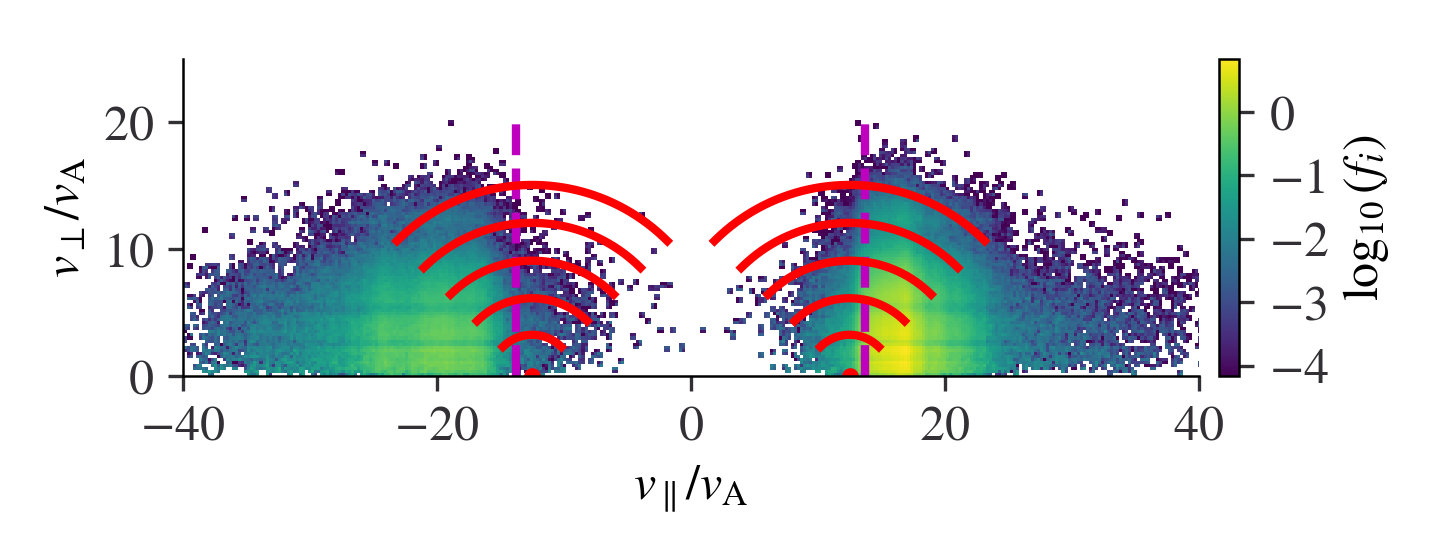}
    \caption{Excitation of magnetosonic waves through anomalous cyclotron resonance with anisotropic ion beams. The ion velocity distribution, color-coded by phase space density, is collected from the spatial region $-2 \leq x/d_i < 2$. Magenta lines denote the velocities $v_r = \pm 13.75 v_\mathrm{A}$ corresponding to the anomalous resonance condition. Red curves centered at the frame of the wave phase velocity, $\omega / (k v_\mathrm{A})$, represent resonance diffusion paths $(v_\parallel - \omega / k)^2 + v_\perp^2 = \mathrm{constant}$, along which ions propagate during wave-particle interactions.}
    \label{fig:anomalous_cyc_res}
\end{figure}

The maximum wave power in Fourier space occurs at $k d_i = \pm 0.8$, $\omega / \omega_{ci} = 10$, and $\omega / (k v_{\mathrm{A}}) = \pm 12.5$ [Figure \ref{fig:bwave-phasespeed}], corresponding to waves near the wake's center. To understand wave amplification mechanisms, we analyze the ion velocity distribution at this location. Resonance between right-hand polarized magnetosonic waves and left-hand gyrating ions requires ions to exceed the wave phase velocity, thereby reversing the wave polarization from the ions' perspective. This interaction is described by the anomalous cyclotron resonance condition $\omega - k v_{\parallel} = -\omega_{ci}$ with $k v_\parallel > 0$, yielding a resonance velocity of $v_{r} = \pm 13.75 v_\mathrm{A}$. Since $\omega \gg \omega_{ci}$ and $k v_\parallel \gg \omega_{ci}$, the anomalous resonance condition approximates the Landau resonance condition $\omega \approx k v_\parallel$. Near $v_r$, the resonant diffusion curves are nearly horizontal ($v_\perp \approx \mathrm{constant}$) [Figure \ref{fig:anomalous_cyc_res}]. Along these curves, energy transfers from ions to waves due to the phase space density gradient $\mathrm{sgn}(k) \cdot \partial f / \partial v_{\parallel} > 0$, where $\mathrm{sgn}(\cdot)$ is the sign function. This anomalous cyclotron resonance may operate continuously as ions stream into the lunar wake. Once ion beam velocities exceed the phase velocities of magnetosonic waves, these waves experience local amplification as they propagate toward the wake's center, resulting in convective growth [Figures \ref{fig:ion_vcl} and \ref{fig:bwave-polarization}(a)].

\subsection{Electromagnetic ion cyclotron waves and normal cyclotron resonance}
Electromagnetic ion cyclotron (EMIC) waves, at frequency below the ion cyclotron frequency, may be simultaneously excited by anisotropic ion beams. For this interaction to occur, ions must counter-propagate with respect to waves to up-shift the wave frequency to the ion gyrofrequency, as described by the normal cyclotron resonance condition $\omega - k v_\parallel = \omega_{ci}$ with $k v_\parallel < 0$. These EMIC waves extract energy from the temperature anisotropy of ion beams. They propagate outward from the wake's center and exhibit much lower frequencies ($\omega < \omega_{ci}$) compared to the observed magnetosonic waves ($\omega \approx 10\, \omega_{ci}$). Our simulation's limited time span of $0 \leq t \omega_{pi} \leq 14400$ (equivalent to $0 \leq t \omega_{ci} \leq 1.8$) is insufficient for EMIC waves to develop fully. Nevertheless, we anticipate that EMIC waves will be excited in the lunar wake, which could be verified through spacecraft observations or more extensive computational simulations.

\subsection{Impact of reduced ion-to-electron mass ratio on electromagnetic instabilities}
In our simulation, electrostatic processes govern the streaming velocities of ion beams. The normalized ion beam velocities $v_\parallel / c_s$ before and after shock formation depend primarily on intrinsic properties of the ion acoustic instability, which remain unchanged as long as $m_i / m_e \gg 1$ [see Sections \ref{subsec:energy-conversion} and \ref{subsec:scaling}]. The generation of magnetosonic and EMIC waves is controlled by ion beam velocities normalized to the Alfvén velocity ($v_\parallel / v_\mathrm{A}$). However, due to our reduced ion-to-electron mass ratio, the $c_s / v_\mathrm{A}$ ratio is approximately 4 times larger than under realistic conditions [see Table \ref{tab:sim_parameters}]. Consequently, with a realistic ion-to-electron mass ratio, we would expect $v_\parallel / v_\mathrm{A} \approx 3.75$, rather than $v_\parallel / v_\mathrm{A} \approx 15$ as observed with $m_i / m_e = 100$ [Figures \ref{fig:ion_vcl} and \ref{fig:anomalous_cyc_res}]. Since anomalous cyclotron resonance requires higher resonance velocity than normal cyclotron resonance \cite{gary1993theory}, EMIC waves (normal cyclotron resonance) may exhibit larger growth rates than magnetosonic waves (anomalous cyclotron resonance) with the smaller $v_\parallel / v_\mathrm{A}$ values expected under realistic conditions. This discrepancy could be resolved using hybrid simulations \cite<e.g.,>[]{travnivcek2005structure}, but is beyond the scope of this study.\add{ Finally, although we use a reduced ion-to-electron mass ratio, this mainly affects the scale separation between ion and electron physics and the $c_s / v_{\mathrm{A}}$ ratio, our simulation domain remains adequate for resolving ion-scale electromagnetic waves with a realistic lunar radius-to-ion inertial length ratio of $R_l / d_i = 13.2$.}

\section{Summary and discussion}\label{sec:summary}
In summary, we study plasma refilling of the lunar wake using a fully kinetic approach, with more realistic ion-to-electron mass ratio $m_i / m_e = 100$ and lunar radius $R_l / d_i = 13.2$ compared to previous simulations of this kind. The main results are summarized as follows.
\begin{enumerate}
    \item In the early stage of plasma refiling, our simulation results agree well with the self-similar solutions of plasma expansion into a vacuum, including the exponential decay of plasma density into the wake, the linear ion flow velocity profile in the self-similar coordinate, and the exponential cooling of ion temperature parallel to the ambient magnetic field [Figure \ref{fig:theory-PIC-comparison}].

    \item Our simulation captures the dynamics of the ion front, where quasi-neutrality breaks down, including the time evolution of the ion front position and the decay of the associated electric field as $1/(t \omega_{pi})$ [Figure \ref{fig:theory-PIC-Efield}].

    \item We demonstrate that the electron temperature in the expansion direction decays exponentially into the wake at the same rate as plasma density, i.e., $T_{e, x} / n_e = \mathrm{constant}$ [Figure \ref{fig:EOS-electron}].

    \item In the later stage of plasma refilling, counter-streaming supersonic ion beams collide in the central wake, generating Debye-scale electrostatic shocks of ion-acoustic nature [Figures \ref{fig:wave-xt-center} and \ref{fig:wave-ps-center}].
    
    \item Across these shocks, ion kinetic energy is predominantly converted to electron thermal energy, with the shock potential jump sculpting flat-top electron velocity distributions [Figure \ref{fig:ps-zoom-ene}].

    \item Due to simultaneous acceleration and cooling parallel to the expansion direction, ions form anisotropic beam distributions [Figure \ref{fig:ion_vcl}] that can drive magnetosonic and EMIC wave instabilities.
\end{enumerate}

These simulation results show potential for validation using spacecraft observations, particularly from ARTEMIS, which has extensively sampled the lunar wake at varying distances from the Moon. We expect to detect counter-streaming, supersonic ion beams and their resulting electrostatic shocks, along with perpendicular anisotropy in these beams and the generation of electromagnetic ion-scale waves. \change{Future computational work should employ two-dimensional PIC simulations to model wave propagation oblique to the magnetic field and plasma entry into the wake perpendicular to the field. These enhanced simulations should also incorporate more realistic conditions, including non-perpendicular solar wind flows relative to interplanetary magnetic fields and non-Maxwellian solar wind electron distributions with halo and strahl components, which would introduce additional complexity to the system.}{Future one-dimensional PIC simulations should incorporate realistic non-Maxwellian solar wind electron distributions, including both halo and strahl components. These non-thermal electron populations significantly modify wake potential structures, which subsequently govern electron transport and acceleration dynamics along magnetic field lines }\cite{halekas2014first}\add{.}

\add{Lunar wake refilling involves several complex processes that require multi-dimensional simulations to capture fully. A key mechanism is the perpendicular entry of plasma into the wake, driven by gyrating ions with finite gyroradii that transport plasma across magnetic field lines }\cite{nishino2009pairwise,nishino2009solar,nishino2010effect}\add{. This cross-field transport creates a distinctive pattern: compressed plasma regions within the lunar wake and corresponding rarefaction zones in the surrounding solar wind }\cite{halekas2005electrons,zhang2014three,fatemi2013lunar}\add{.}

\add{This perpendicular refilling process exhibits striking similarities to phenomena observed in other plasma environments. These include the solar wind-barium cloud interactions studied in the Active Magnetospheric Particle Tracer Explorers (AMPTE) experiment }\cite{papadopoulos1987collisionless,bernhardt1987observations}\add{, plasma expansion dynamics in laser-laboratory experiments }\cite{bonde2015electrostatic,bondarenko2017collisionless}\add{, and foreshock ion interactions with magnetic discontinuities that generate foreshock transients }\cite{an2020formation,liu2020magnetospheric}\add{. The lunar wake thus serves as a natural laboratory for studying cross-field plasma transport across diverse parameter regimes, enabling comparative analysis of fundamentally similar physical processes.}

\add{Multi-dimensional simulations are also essential for capturing the fluid and kinetic instabilities that develop during perpendicular refilling of the lunar wake. Pressure gradients associated with perpendicular plasma entry can drive flute instabilities, leading to interchange motions between plasma-filled and plasma-depleted flux tubes }\cite{borisov2000plasma}\add{. These same pressure gradients may also trigger shorter-wavelength drift instabilities that further modify the refilling dynamics }\cite{roussos2012energetic}\add{. Understanding the linear stability conditions and nonlinear evolution of these instabilities in the lunar wake environment represents an important avenue for future research. Quantifying their relative contributions to cross-field transport would provide crucial insights into the efficiency and timescales of wake refilling processes.}

\section*{Open Research Section}
The simulation data and Jupyter notebooks used in the study are available at Dryad via \citeA{An2025plasma} [\url{https://doi.org/10.5061/dryad.8w9ghx3zv}] with the license CC0 1.0 Universal [\url{https://creativecommons.org/publicdomain/zero/1.0/}].

\acknowledgments
This work was supported by NASA contract NAS5-02099 and NASA grants NO.~80NSSC22K1634 and No.~80NSSC23K0086. We would like to acknowledge high-performance computing support from Derecho (\url{https://doi.org/10.5065/qx9a-pg09}) provided by NCAR's Computational and Information Systems Laboratory, sponsored by the National Science Foundation \cite{derecho}.

%
%


%
%
%
%
%

\end{document}